\documentclass[aps,prd,twocolumn,superscriptaddress,groupedaddress]{revtex4-1}  % for review and submission
\usepackage{graphicx}  % needed for figures
\usepackage{dcolumn}   % needed for some tables
\usepackage{bm}        % for math
\usepackage{amssymb}   % for math
\usepackage{amsmath}
\usepackage{color}
\usepackage{acronym}
\usepackage{algcompatible}
\usepackage{newfloat}
\usepackage{cancel}
\usepackage{subcaption}
\graphicspath{{./figures/}}

\newcommand{\dcc}{LIGO-P1900078}

\DeclareMathOperator*{\argmax}{argmax}

\DeclareFloatingEnvironment[
    fileext=loa,
    listname=List of Algorithms,
    name=ALGORITHM,
    placement=tbhp,
]{algorithm}

\begin{document}

\title{SOAP: A generalised application of the Viterbi algorithm to searches for continuous gravitational-wave signals}

\author{Joe Bayley}
\author{Chris Messenger}
\author{Graham Woan}
\affiliation{Institute of Gravitational research, University of Glasgow}

\def\commitID{commitID: 1b9b3606f86fdf4fa2498843cc6334a4ff42f895}
\def\commitDATE{Mon Mar 11 17:24:58 2019 +0000}

\date{\commitDATE\\\mbox{\small \commitID}\\\mbox{\dcc}}

\begin{abstract}
All-sky and wide parameter space searches for continuous gravitational waves are generally  template-matching schemes which test a bank of signal waveforms against data from a gravitational wave detector.  Such searches can offer optimal sensitivity for a given computing cost and signal model, but are highly-tuned to specific signal types and are computationally expensive, even for semi-coherent searches. We have developed a search method based on the well-known Viterbi algorithm which is model-agnostic and has a computational cost several orders of magnitude lower than template methods, with a modest reduction in sensitivity. In particular, this method can search for signals which have an unknown frequency evolution. We test the algorithm on three simulated and real data sets: gapless Gaussian noise, Gaussian noise with gaps and real data from the final run of initial LIGO (S6). We show that at 95\% efficiency, with a 1\% false alarm rate, the algorithm has a depth sensitivity of $\sim 33$, $10$ and $13$\,Hz$^{-1/2}$ with corresponding SNRs of $\sim 60$, $72$ and $74$ in these datasets. we discuss the use of this algorithm for detecting a wide range of quasi-monochromatic gravitational wave signals and instrumental lines.
\end{abstract}

\maketitle

% define some acronyms
\acrodef{GW}[GW]{gravitational-wave}
\acrodef{CW}[CW]{continuous wave}
\acrodef{NS}[NS]{neutron star}
\acrodef{EM}[EM]{electromagnetic}
\acrodef{SNR}[SNR]{signal-to-noise-ratio}
\acrodef{LIGO}[LIGO]{Laser Interferometer Gravitational-wave Observatory}
\acrodef{SFT}[SFT]{short Fourier transform}
\acrodef{UCD}[UCD]{up, centre or down}
\acrodef{MDC}[MDC]{mock data challenge}
\acrodef{PSD}[PSD]{power spectral density}
\acrodef{ROC}[ROC]{receiver operating characteristic}
\acrodef{RMS}[RMS]{root median square}
\acrodef{MCMC}[MCMC]{Markov-Chain Monte Carlo}

%%%%%%%%%%%%%%%%%%%%%%%%%%%%%%%%%%%%%%%%%%%%%%%%%%%%%%%%%%%%%%%%%%%%%%%%%
%%%%%%%%%%%%%%%%%%%%%%%%%%%%%%%%%%%%%%%%%%%%%%%%%%%%%%%%%%%%%%%%%%%%%%%%%
%%%%%%%%%%%%%%%%%%%%%%%%%%%%%%%%%%%%%%%%%%%%%%%%%%%%%%%%%%%%%%%%%%%%%%%%%
\section{\label{intro:1}Introduction}
%%%%%%%%%%%%%%%%%%%%%%%%%%%%%%%%%%%%%%%%%%%%%%%%%%%%%%%%%%%%%%%%%%%%%%%%%
%
% introduce the signal
%
One of the main targets for current ground based \ac{GW} detectors, including \ac{LIGO}~\cite{LIGO, LIGO2015} and Virgo~\cite{VIRGO, Acernese_2014}, are sources of continuous gravitational waves. These are long-duration, quasi-monochromatic  sinusoidal signals that are well-modelled by a Taylor series expansion in the signal phase. A likely source of such signals are rapidly spinning non axisymmetric neutron stars.  A number of possible emission mechanisms are outlined in~\cite{Prix2009,Owen:2009fg}.

%
% introduce semi-coherent search algorithms
%
These types of \acp{GW} are expected to give strain amplitudes that are significantly below the detector's noise spectral density, and need sensitive search algorithms for detection. The most sensitive method is to use a coherent matched filter which requires knowledge of the waveform beforehand such that it can be coherently correlated with the data. This approach is used in searches for gravitational signals from known pulsars such as~\cite{Dupuis2005,Astone_2010,PhysRevD.58.063001,O1knownpulsar2017,O2knownpulsar:2019}. For broad parameter space searches, where the parameters of the signal are unknown, a large number of template waveforms must be used to sufficiently cover the parameter space.  This approach rapidly becomes computationally impractical as the search space grows, so semi-coherent search methods have been developed to deliver the maximum overall sensitivity for a given computational cost. Semi-coherent searches break the data up into sections of either time or frequency and perform a coherent analysis on these sections separately. These intermediate results can then be recombined incoherently in a number of different ways to form the final search result outlined in ~\cite{PhysRevD.61.082001,Pisarski:2019vxw} and references therein.

%
% introduce SOAP
%
The analysis that we present here is known as SOAP \cite{soap} and is based on the Viterbi algorithm~\cite{Viterbi1967}. The algorithm models a process that has a discrete number of states at discrete time steps, and computes the set of states which gives the highest probability (suitably defined) given the data. Our implementation of SOAP is intended as a stand-alone search which is naturally non-parametric and has broad applications to both searches for known signal types and signals which have an unknown frequency evolution. The algorithm works in time-frequency plane, where our `states' are represented by the time and frequency coordinates of a potential signal. We can then find the most probable set of frequencies a possible signal could have, i.e. we can find the most probable track in frequency as a function of time. This is not the first application of the Viterbi algorithm to \ac{GW} data. Another variant of the algorithm \cite{Suvorova2016} has recently been used, amongst other applications, as part of a \ac{CW} search to track a pulsar with randomly wandering spin frequency~\cite{PhysRevD.97.043013, PhysRevD.96.102006,PhysRevD.95.122003, Abbott:2018hgk, 2018arXiv181003577S}.

%
% describe the paper structure
%
In the next section we will describe the Viterbi algorithm and the basic SOAP implementation to \ac{GW} time-frequency data. We then describe additional features to the algorithm, including the use of data from multiple detectors. As well as this we describe methods used to ignore instrumental effects in the data, such as incoherently summing data and a `line aware' statistic. In the final section as well as a test of the computational cost of the search, we show results of a search performed on datasets of increasing complexity: Gaussian noise with no gaps (i.e., contiguous in time), Gaussian noise with gaps simulating real data more accurately, and finally real \ac{LIGO} data taken during the sixth science run. 

%%%%%%%%%%%%%%%%%%%%%%%%%%%%%%%%%%%%%%%%%%%%%%%%%%%%%%%%%%%%%%%%%%%%%%%%%
%%%%%%%%%%%%%%%%%%%%%%%%%%%%%%%%%%%%%%%%%%%%%%%%%%%%%%%%%%%%%%%%%%%%%%%%%
\section{\label{viterbi}The Viterbi Algorithm}
%%%%%%%%%%%%%%%%%%%%%%%%%%%%%%%%%%%%%%%%%%%%%%%%%%%%%%%%%%%%%%%%%%%%%%%%%
%%%%%%%%%%%%%%%%%%
%
% Into to viterbi
%
The Viterbi algorithm is an efficient method for determining the most probable set of states (a single `track' of steps on the time-frequency plane) in a Markov model dependent on data, where the model has a discrete number of states at each step. Rather than computing the probability of every possible track and selecting the most probable, the algorithm maximises this probability after every discrete step. As a result, a partial track which cannot ultimately be the most probable is rejected before the next step is calculated, and only a fraction of all possible tracks need to be computed to find the one that is most probable.

%
% Defining varibles and what data we use
%
In this work we apply the Viterbi algorithm to a \ac{GW} strain time-series to find the most probable track of a single variable-frequency signal in the noisy data.  We divide the time series into $N$ equal-length and contiguous segments ${\bm x}_j$,  defining the set $D \equiv \{{\bm x}_j\}$. The `states' in the model correspond to the frequencies a signal could have in each segment. A `track' is a list of such frequencies ${\bm \nu}\equiv \{\nu_j\}$, where  $\nu_j$ is the frequency in the segment ${\bm x_j}$.

%
% defining probabilities
%
Our objective is to calculate the most probable track given the data, i.e., the
track that maximises $p({\bm \nu}\mid D)$. Using Bayes theorem, this posterior probability can
be written as
\begin{equation}
\label{viterbi:bayes}
  p({\bm \nu} \mid D) = \frac{p({\bm \nu})p(D \mid {\bm \nu})}{p(D)},
\end{equation}
where $p({\bm \nu}) $ is the prior probability of the
track, $p(D \mid{\bm \nu})$ is the likelihood of the track (i.e., the
probability of the data given the track) and $p(D)$ is the model evidence (or
marginalised likelihood).

The Viterbi algorithm treats the track as the result of a Markovian process,
such that the current state depends only on the previous state. It is
therefore useful to split the track's prior into a set of transition
probabilities such that
\begin{align}
\label{viterbi:prior}
p({\bm \nu}) &= p(\nu_{N - 1}, \ldots, \nu_1, \nu_0)\nonumber \\
%&=  p(\omega_{N} \mid \omega_{N-1}, \ldots, \omega_1,\omega_0)p(\omega_{N-1} \mid \omega_{N-2}, \ldots, \omega_1,\omega_0) \ldots p(\omega_1 \mid \omega_0)p(\omega_0) \\
&= p(\nu_{N - 1} \mid \nu_{N-2})p(\nu_{N-2} \mid \nu_{N-3}) \dots p(\nu_1 \mid \nu_0)p(\nu_0) \nonumber \\
&= p(\nu_0)\prod_{j=1}^{N-1}p(\nu_{j} \mid \nu_{j-1}),
\end{align}
where $p(\nu_0)$ is the prior probability that the signal in the first time
step has a frequency $\nu_0$ and $p(\nu_{j} \mid \nu_{j-1})$ is the
prior `transition' probability for $\nu_j$ given the frequency at the last
step was $\nu_{j-1}$.

The noise in each of the segments can be treated as independent, so the
likelihood component in Eq.~\ref{viterbi:bayes} can be factorised as
\begin{equation}
\label{viterbi:likelihood}
p(D \mid {\bm \nu}) = \prod_{j=0}^{N-1}p({\bm x_j} \mid \nu_j),
\end{equation}
 where $p({\bm x_j} \mid \nu_j)$ is the likelihood of our
signal having a frequency $\nu_j$ in the $j$th segment.

Using Eqs.~\ref{viterbi:bayes},\ref{viterbi:prior} and
\ref{viterbi:likelihood}, the posterior probability is then
\begin{equation}
\label{viterbi:posterior}
    p({\bm \nu} | D) =
    \frac{p(\nu_0)p({\bm x_0} | \nu_0) \displaystyle\prod_{j=1}^{N-1}p(\nu_{j}
| \nu_{j-1})p({\bm x_j} | \nu_j)}{\displaystyle\sum_{S}
\left\{p(\nu_0)p({\bm x_0} | \nu_0)\displaystyle\prod_{j=1}^{N-1}p(\nu_{j} |
\nu_{j-1})p({\bm x_j} | \nu_j)\right\}} ,
\end{equation}
where in the denominator we must sum over all possible tracks
$S$. We require the specific track, or set of frequencies, $\hat{\bm
\nu}$ that  maximises the posterior probability, and which therefore
maximises the numerator  on the right-hand side of  Eq.~\ref{viterbi:posterior}, i.e.,
\begin{equation}
\label{viterbi:maxtrack}
  p(\hat{\bm \nu} | D) \propto \max_{\bm \nu}{\left[p(\nu_0)p({\bm x_0} |
\nu_0) \prod_{j=1}^{N-1}p(\nu_{j} |\nu_{j-1})p({\bm x_j} | \nu_j)\right]}.
\end{equation}
This track also maximises the log of the probability,
\begin{equation}
\label{viterbi:maxtracklog}
\begin{split}
  \log p(\hat{\bm \nu} | D)  = \max_{{\bm \nu}}{\biggl\{ \log p(\nu_0) + \log p({\bm x_0} | \nu_0)  } \\
 \left. \sum_{j=1}^{N-1} \biggl[ \log p(\nu_{j} | \nu_{j-1}) + \log p({\bm x_j}
| \nu_j) \biggr] \right\} + \text{const}.
  \end{split}
\end{equation}
The Viterbi algorithm finds the most probable track $\hat{\bm \nu}$ by calculating the quantities in Eq,~\ref{viterbi:maxtracklog} for each frequency at each time step. In the following sections we explain how this is achieved in practice.

%%%%%%%%%%%%%%%%%%%%%%%%%%%%%%%%%%%%%%%%%%%%%%%%%%%%%%%%%%%%%%%%%%%%%%%%%
%%%%%%%%%%%%%%%%%%%%%%%%%%%%%%%%%%%%%%%%%%%%%%%%%%%%%%%%%%%%%%%%%%%%%%%%%
\subsection{\label{viterbi:transition}The transition matrix}
%%%%%%%%%%%%%%%%%%%%%%%%%%%%%%%%%%%%%%%%%%%%%%%%%%%%%%%%%%%%%%%%%%%%%%%%%
%
% define the transition matrix
%
We define the `transition matrix' $T$ as the matrix that stores the prior log-probabilities $\log p(\nu_j \mid \nu_{j-1})$. These transition probabilities depend only on the size and direction of the transition, and in our case correspond to a jump in frequency when moving from the $(j-1)$th to the $j$th state. It is within the transition matrix that we impose some loose model constraints. For example it is usual in the time-frequency plane for frequencies to only have discrete values (frequency bins) and a track might only be allowed to move by one bin in each time step, restricting it to a \ac{UCD} transition or `jump' or equivalently setting the size of the first dimension of the transition matrix $n_1 = 3$. We can also impose that the transition probabilities are independent of the current track location in frequency, i.e. $p(\nu_j \mid \nu_{j-1})=p(\nu_{j+k} \mid \nu_{j+k-1})$. This leads to the transition matrix containing only three numbers, corresponding to the three prior log-probabilities that the track was in the corresponding \ac{UCD} frequency bin at the previous time step. These numbers are chosen to reflect the prior probability of a frequency deviation in the track and depend on the class of signals that one wishes to detect.

In later sections we will consider more complex situations in which the transition matrix describes the prior probability associated with sequences of even earlier transitions (`memory') and the case where there are multiple detectors. In these cases the number of dimensions of the transition matrix can grow substantially to account for the extra complexity of the problem.

%%%%%%%%%%%%%%%%%%%%%%%%%%%%%%%%%%%%%%%%%%%%%%%%%%%%%%%%%%%%%%%%%%%%%%%%%
%%%%%%%%%%%%%%%%%%%%%%%%%%%%%%%%%%%%%%%%%%%%%%%%%%%%%%%%%%%%%%%%%%%%%%%%%
\subsection{\label{viterbi:single}Single detector}
%%%%%%%%%%%%%%%%%%%%%%%%%%%%%%%%%%%%%%%%%%%%%%%%%%%%%%%%%%%%%%%%%%%%%%%%%
%
% single detector algorithm,
%
We will first consider the simple case of a single dataset $D$, generated by a single gravitational wave detector, and consider only a one-dimensional transition matrix, i.e. the transition matrix contains only three numbers refering the the track `jumps'. We will make use of discrete Fourier transforms so that frequencies, and hence the track frequencies, are also discrete. These frequencies will be indexed by $k$ and therefore $\nu_j \rightarrow \nu_{j,k}=k(j)\Delta f$ where $\Delta f=1/T$ is the frequency bin width for a segment of duration $T$.

 The Viterbi algorithm determines the most probable track on the time-frequency plane by calculating the value of Eq.~\ref{viterbi:maxtracklog} for every discrete Fourier frequency, incrementally in time. In other words, at each time segment it finds the most probable earlier track which ends at each particular frequency. On reaching the final segment it can look back to identify the most probable track connecting segment 1 to segment $N$.

There are two main components to Eq.~\ref{viterbi:maxtracklog}: the transition probabilities $p(\nu_j \mid \nu_{j-1})$ and the likelihoods $p({\bm x_j} \mid \nu_j)$. The transition probabilities are pre-calculated and stored in a transition matrix according to Sec.~\ref{viterbi:transition} above, and details of how these are calcuated are described in Sec.~\ref{results}. To calculate the likelihood we follow the approach of~\cite{Bretthorst1988} which gives, under the assumption of a single sinusoidal signal in additive Gaussian noise in data segment ${\bm x_j}$,
\begin{equation}
\label{viterbi:single:likelihood}
p({\bm x_j} \mid \nu_{j,k}, \sigma_{j,k}, I) \propto
\exp\left[{C(\nu_{j,k})}\right].
\end{equation}
where
$C_{j,k}(\nu_{j,k})$ is the Schuster periodogram normalised to the noise variance at
frequency $\nu_{j,k}$ of segment $j$. This is equivalent to the log-likelihood, and is defined as
\begin{equation}
\label{viterbi:periodogram}
C(\nu_{j,k}) \equiv C_{j,k}=  \frac{1}{\sigma_{j,k}^2}\frac{1}{N_{\rm s}}\left|\sum_{r=0}^{N_{\rm
s}-1}x_{j,r} {\rm
e}^{i\nu_{j,k} t_r}\right|^2,
\end{equation}
where $N_{\rm s}$ is the number of data points in each segment and $t_{r}$ is the time corresponding to $x_{j,r}$, the $r$th sample in the $j$th data segment. $\sigma_{j,k}^2$ is the noise variance and is calculated as an estimate of the noise \ac{PSD} in the $k$th sample and the $j$th data segment.
The log-likelihoods of each segment can be calculated at discrete frequencies before running the algorithm by computing the power spectra for each segment from discrete Fourier transforms of the data. In the \ac{GW} field these standard data forms are known as \acp{SFT}.
%
% Using odds ratios instead of likelihood
%
It is worth noting at this point that it is also possible to write this as a likelihood ratio, and therefore write out detection statistic as a log-odds ratio, however, we will discuss this in more depth in Sec.~\ref{viterbi:las}. 

\begin{figure}
\centering
\begin{subfigure}[h]{\columnwidth}
\includegraphics[width=\columnwidth]{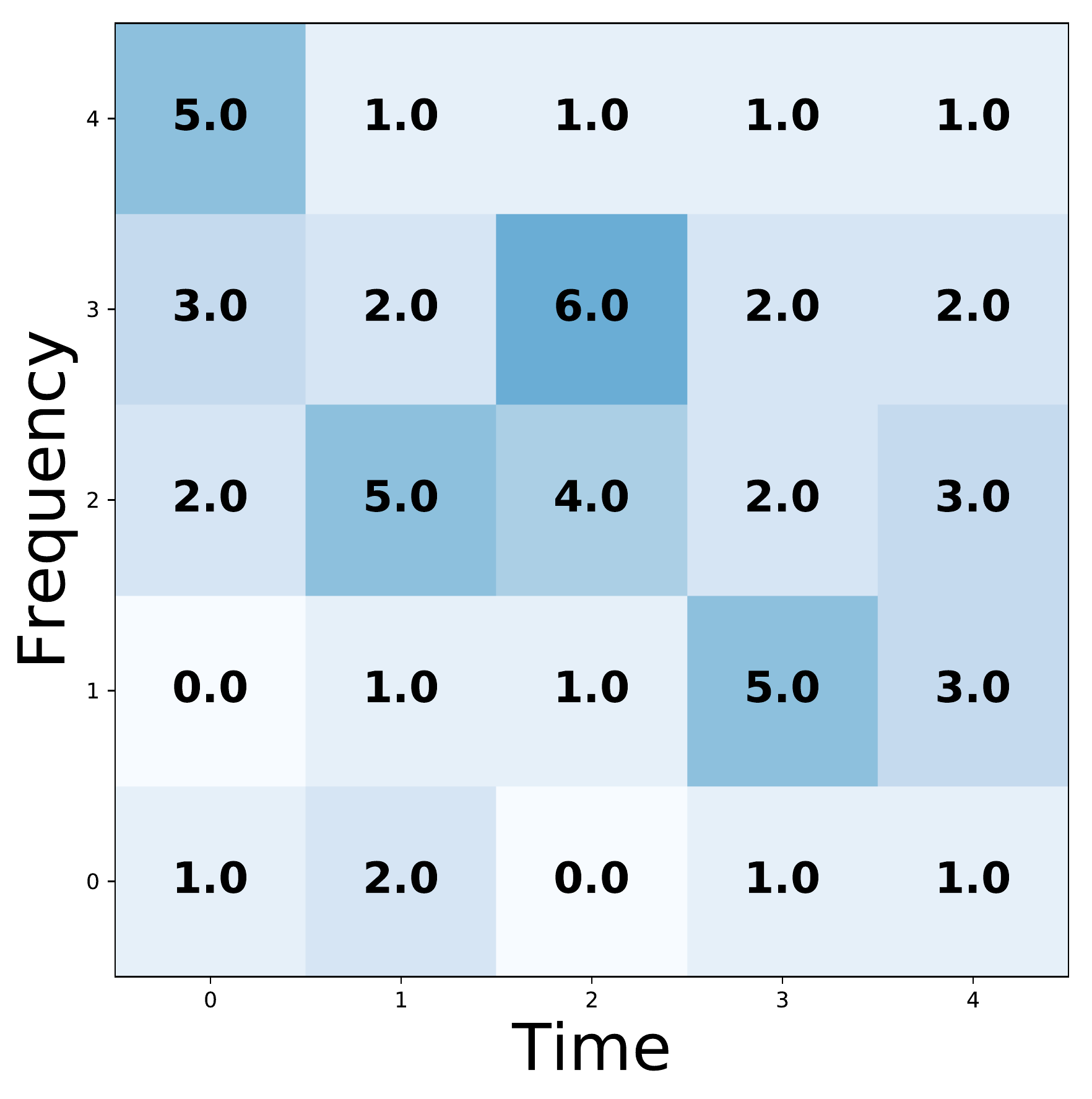}
\caption{The input data}
\label{viterbi:plot:data}
\end{subfigure}

\begin{subfigure}[h]{\columnwidth}
\includegraphics[width=\columnwidth]{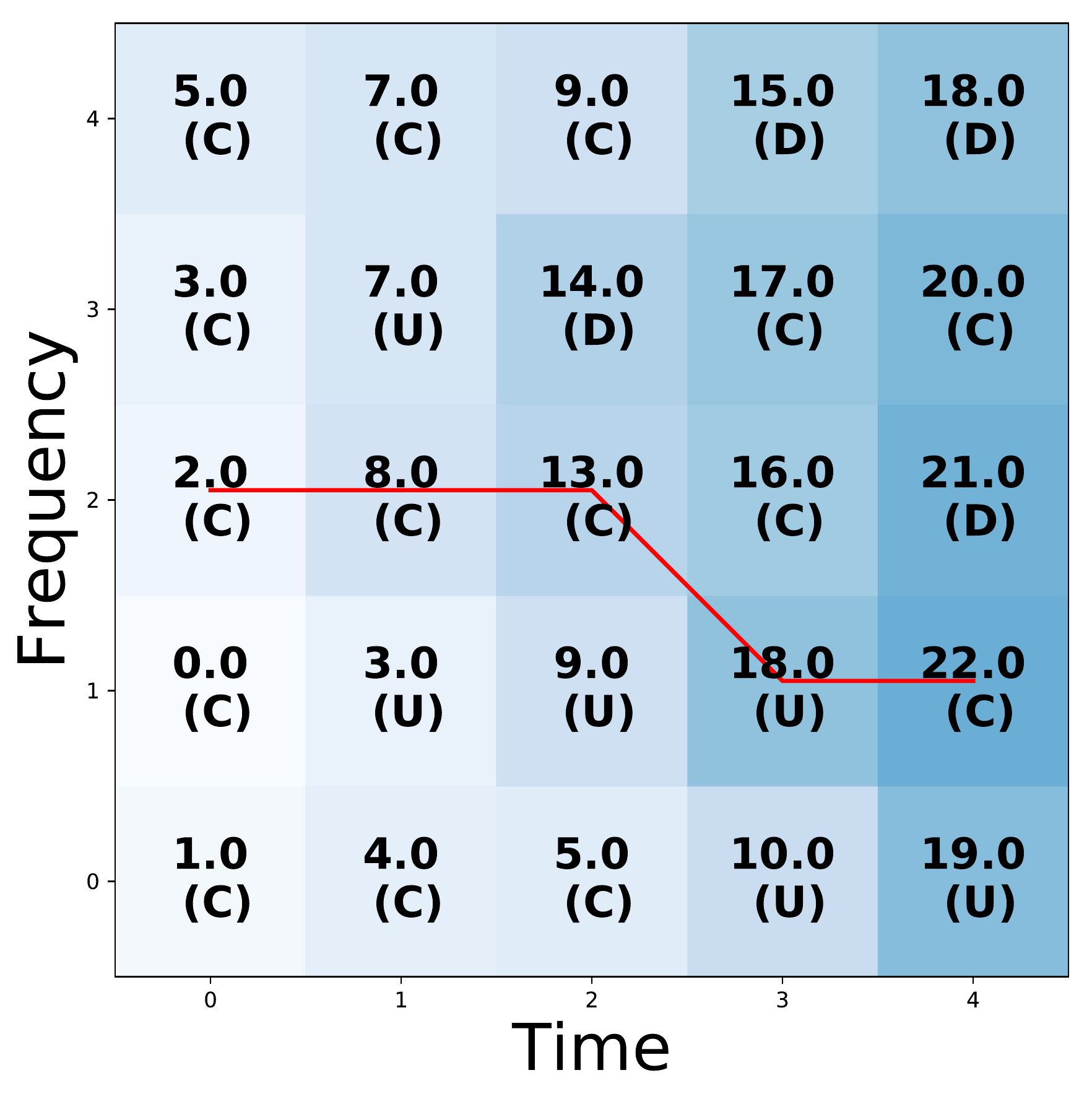}
\caption{The log-probabilities, jumps, and most probable path}
\label{viterbi:plot:likelihood}
\end{subfigure}

\caption{ Fig.~\ref{viterbi:plot:data} shows
the observed data, i.e the log-likelihood values $C_{j,k}$. Fig.~\ref{viterbi:plot:likelihood} shows the calculated
log-probabilities $V_{j,k}$. $A_{j,k}$ is shown in parentheses, where the \ac{UCD}
components correspond to $i= [-1,0,1]$ respectively. The red line shows the
path that gives the maximum probability. The transition matrix for the \ac{UCD} jumps is $[0,1,0]$ and corresponds to the un-normalised prior
log-probabilities of these jumps occurring.}
\label{viterbi:plots}
\end{figure}

The Viterbi algorithm records two quantities for each frequency and time bin: The first, $V_{j,k}$, contains the value defined by Eq.~\ref{viterbi:maxtracklog}, which is the log-probability of the most probable path ending in position $j,k$. The second, $A_{j,k}$, is the transition, or `jump', used to achieve the most probable path. The algorithm can be divided into three main sections: initialisation, iteration and identification. These three sections are described in pseudo-code in Alg.~\ref{viterbi:single:algorithm} and a simple demonstration of the algorithm at work is shown in Fig.~\ref{viterbi:plots}.
%
%
%   pseudo algorithm
%
\begin{algorithm}
\begin{algorithmic}[1]
\STATE{Input: ${C}$, $T$} \COMMENT{log-likelihood,transition matrix}
\STATE{Output: $\hat{\bm \nu}$, $V$, $A$} \COMMENT{most probable track, track probabilities, jumps}
\STATE
\STATE{ {\it Initialisation}}
\FOR{Frequency ($\nu_{0,k}$), $k=0 \rightarrow M-1$}
    \STATE{$V_{0,k} =  { C_{0k}} $ }
    \STATE{$A_{0,k} = 0$}
\ENDFOR
\STATE
\STATE{ {\it Iteration}}
\FOR{Segment, $j=0 \rightarrow N-1$}
    \FOR{Frequency ($\nu_{j,k}$), $k=0 \rightarrow M-1$}
        \STATE{$V_{j,k} = {\max\limits_{i }  ({ C_{j,k}} + T_i + V_{j-1,j+i})}$}
        \STATE{$A_{j,k} = {\argmax\limits_{ i }  ({ C_{j,k}} + T_i + V_{j-1,j+i})}$}
    \ENDFOR
\ENDFOR
\STATE
\STATE{ {\it Identification}}
\STATE{$\hat{\nu}_{N-1} = \argmax_k (V_{N-1,k})$}
\FOR{Segment, $j=N-1 \rightarrow 0$}
	\STATE{$\hat{\nu}_j = \hat{\nu}_{j+1} + A_{j,\nu_{k+1}}$}
\ENDFOR
\end{algorithmic}
\caption{The Viterbi algorithm in pseudo-code. $N$ is the number of segments, $M$ is the number of frequency bins in each segment. Here the maximisations over $i$ run between $\pm (n_1-1)/2$ where $n_1$ is the size of the transition matrix. The values from Eq.~\ref{viterbi:maxtracklog} are stored in $V$, and the jumps are stored in $A$. The most probable track is denoted by $\hat{\bm \nu}$.\label{viterbi:single:algorithm}}
\end{algorithm}
%
%
% example plot to work through
%
%

\begin{description}
%
% Initialisation
%
\item[Initialisation] The two parts of Eq.~\ref{viterbi:maxtracklog},  $\log p(\nu_0)$ and $\log p({\bm x_0} \mid \nu_0)$, must be computed before the main recursive part of the algorithm can start. Therefore, the initialisation section (lines 5--8) in Alg.~\ref{viterbi:single:algorithm} calculates the first column in the lower panel of Fig.~\ref{viterbi:plots}. A priori, there is no preferred initial frequency, so we take the log-prior $\log p(\nu_{0,k})$ to be uniform over the complete frequency range. As a result, this is does not affect the maximisation for any jump, therefore, can be omitted from the calculation. We then use the pre-calculated log-likelihood values $C_{0,k}$ to fill the track probabilities $V_{0,k}$.  There is no previous position to jump from in this case, so the transition probabilities are irrelevant and $A_{0,k}$ are set to zero.
%
% Iteration
%
\item[Iteration] The main part of the calculation is the sum in Eq.~\ref{viterbi:maxtracklog}. Lines 11--16 in Alg.~\ref{viterbi:single:algorithm} calculate the most probable tracks that end at each frequency bin for each segment by using
    \begin{equation} \label{viterbi:single:vitsum}
    V_{j,k} = \max_{i}({C_{j,k} }+ T_{i} + V_{j-1,k+i}),
    \end{equation}
    where $i$ is the size and direction of the jump. For example, in Fig.~\ref{viterbi:plots} columns 1--4 are calculated in order using Eq.~\ref{viterbi:single:vitsum}, where it maximises over three possible previous positions in frequency. These positions are the frequency bins \ac{UCD} of the current position. The size and direction of the jump, $i$, which gives the maximum probability is then saved to $A_{j,k}$. These are shown in parentheses below the log-probabilities in Fig.~\ref{viterbi:plots} where \ac{UCD} correspond to values of $i = [-1,0,1]$ respectively.
%
% Identification
%
\item[Identification] The final stage of the algorithm identifies the most probable track. This is done by initially finding the highest log-probability values in the final time segment, $\max_k(V_{N-1,k})$ (line 19 in Alg.~\ref{viterbi:single:algorithm}). In the lower panel of Fig.~\ref{viterbi:plots} this is located at position $j,k = 4,1$ with $V_{4,1} = 22$. To find the track which corresponds to this, the values in $A_{jk}$ are followed backwards from this position (lines 20--21). For example, in Fig.~\ref{viterbi:plots} the final position is $j,k = 4,1$ and $A_{j,k} = \rm{Center} = 0$, this means that at the previous segment the most probable track was at position $j,k = 4-1,1+0 = 3,1$. At this time $A_{3,1} = R = 1$, therefore, the next track element is at $j,k = 3-1,1+1 = 2,2$. This then continues until $j=0$ whereupon these retraced positions constitute the most probable track, highlighted in red in Fig.~\ref{viterbi:plots}.
\end{description}

%
% mention the limitations - only returns max track.
%
The most probable track is the one traced backwards from the highest probability final segment frequency position. However, tracks can also be traced back from any of the end-frequency positions, returning the most probable track conditional on a given final position. Such tracks should not be confused with the being equal to the second, third, fourth, etc. most probable tracks. Information regarding the rankings and properties of all possible tracks (excluding the most probable and conditionally most probable tracks) is lost during the maximisation procedures computed at each stage in the algorithm --  a necessary consequence of the algorithm's speed and efficiency.

%%%%%%%%%%%%%%%%%%%%%%%%%%%%%%%%%%%%%%%%%%%%%%%%%%%%%%%%%%%%%%%%%%%%%%%%%
%%%%%%%%%%%%%%%%%%%%%%%%%%%%%%%%%%%%%%%%%%%%%%%%%%%%%%%%%%%%%%%%%%%%%%%%%
\subsection{\label{viterbi:multidet}Multiple detectors}
%%%%%%%%%%%%%%%%%%%%%%%%%%%%%%%%%%%%%%%%%%%%%%%%%%%%%%%%%%%%%%%%%%%%%%%%%
%
% introduce what multidetector means
%
If there are $Q$ detectors operating simultaneously we have $Q$ sets of data which can be combined appropriately to provide input to the Viterbi search described above. We must also modify the allowed transitions encoded within the transition matrix to take account of the extra prior constraints that are now available.

%
% Doppler shift at each detector
%
The received instantaneous frequency of a given astrophysical signal will be nearly the same for all ground-based \ac{GW} detectors, and our
algorithm should be sensitive to tracks that show this consistency in
frequency. However there \emph{will} be small differences between the frequencies measured at detectors that are not co-located, due to differential Doppler shifts caused by Earth rotation. As a result the signal could fall in different frequency bins at each detector.

%
% Continue algorithm explanation
%
To account for these small differences in signal tracks in each detector, we reference the observed tracks to a third (pseudo) detector located at the centre of the Earth which would be insensitive to Earth spin. The signal frequencies in each real detector are then allowed to vary within a certain number of frequency bins from the track in the reference detector. In the examples that follow, we only consider the possibilities that the track in each real detector is no more that one frequency bin away from the reference track. We can tune the length of the \acp{SFT} to ensure this is a valid assumption, this is explained in greater detail in \ref{results}.
As well as differences in signal frequency, due to antenna patterns and other effects, the measured signal amplitude may differ between the detectors \cite{Jaranowski1998}. In the following example we assume that the signal has the same amplitude in each detector, however, in Sec.~\ref{viterbi:las} we discuss the case where they differ.

We will now show how the algorithm in Sec.~\ref{viterbi:single} can be modified to handle a two-detector network (i.e., $Q=2$),  however any number of detectors can easily be accommodated. In the two detector case the joint probability of two (real) tracks, $\nu^{(1)}$ and $\nu^{(2)}$, and the geocentric track $\nu$, given the data, is
\begin{equation}
\begin{split}
p(\nu,\nu^{(1)},\nu^{(2)} | D^{(1)},D^{(2)}) \propto p(\nu)p(\nu^{(1)},\nu^{(2)} | \nu) \\
p(D^{(1)} | \nu^{(1)})p(D^{(2)} | \nu^{(2)}),
\end{split}
\end{equation}
where $D^{(1)}$ and $D^{(2)}$ represent the data from the two detectors. The
main difference between this and that described in Sec.~\ref{viterbi:single} is
that the track probabilities $V_{j,k}$ are stored for the geocentric
pseudo-detector. The main iterative calculation (defined for the single
detector case in Eq.~\ref{viterbi:single:vitsum}) now becomes
\begin{equation}
\label{viterbi:multidet:vitsum}
  V_{j,k} = \max_{i,l,m}({C}^{(1)}_{j,k+l} + {C}^{(2)}_{j,k+m} + T_{i,l,m} +V_{j-1,k+i}),
\end{equation}
where ${C}^{(1)}$ and ${C}^{(2)}$ refer to the log-likelihoods in detectors 1 and 2 respectively and the transition matrix $T$ is an $n_1\times n_2 \times n_3$ matrix, where $n_1$ dimension refers to the jump from the previous time step, $n_2$ and $n_3$ refer to the relative frequency positions in each real detector. The transition matrix is now three-dimensional and holds the prior log-probabilities of $p(\nu)$ and $p(\nu^{(1)},\nu^{(2)} | \nu)$.  We now need to maximise over three indices: $i,l$ and $m$. The index $i$ refers to the size and direction of the jump at the geocentre (as before). The indices $l$ and $m$ refer to the number of frequency bins by which the two real tracks deviate from the geocentre track. For example, if the most probable track in the geocentred detector is in bin $j,k = 5,12$ and the values of $i,l,m = 0,-1,1$, then detector 1 is in position $j,k={5,11}$ and detector 2 is in position $j,k={5,13}$ and the geocentred track was in the position $j,k={4,12}$ at the previous time step. As a result, the track at the geocentre is only affected by Doppler modulations from the Earth's orbit whereas the tracks in the real detectors include Doppler modulations from the Earth's spin.

%
% More details on the doppler constraints
%
At every time step the frequency bin position for each real detector is forced to be within $n_l$ or $n_m$ bins of the track in the geocentred detector, where $n_l$ and $n_m$ depend on how much each detector could possibly be Doppler shifted. As mentioned previously, we only consider the case where $n_l=1$ and $n_m = 1$,  allowing the track from each real detector to be at most one frequency bin away from the geocentred track position. While we tune the \ac{SFT} length to keep this condition for different frequencies, it is also possible to tune the values of $n_l$ and $n_m$ to get a similar effect.
%
% how do we do this in practice?
%
The implementation of the multi-detector algorithm is similar to the single detector case described in Sec.~\ref{viterbi:single}.  However in the single detector case there is only a single variable to be maximised over for each time-frequency bin. This variable is the frequency jump from the position in the previous segment. For the multi-detector case there are at least three variables to be maximised over: the probability of the jump, $i$, at the geo-centre and the probability of the signal being in the surrounding positions in each on $Q$ real detectors, $l,m,\dots$. The values of $i,l,m, \dots$ are then saved to $A_{j,k}$ and are ultimately used to reconstruct the most probable consistent tracks in each real detector.

%
%  algorithm explanation
%
As in Sec.~\ref{viterbi:single}, there are three main sections: Initialisation, iteration, and the identification. For the multi-detector case each element is modified as follows.

\begin{description}
% first step calculation
\item[Initialisation] The first-row calculation (lines 5--8) in Alg.~\ref{viterbi:single:algorithm}, are now modified to additionally maximise over the real detector track positions $l$ and $m$. For each time-frequency bin the maximum sum of the log-likelihoods is saved together with the frequency locations of the corresponding tracks in the real detectors. The index $i=0$ is kept constant as there is no previous position.

% all other steps calculation
\item[Iteration] To process the subsequent time segments, lines 13--14 in Alg.~\ref{viterbi:single:algorithm} are modified to account for two (or more) detectors. Line 13 of Alg.~\ref{viterbi:single:algorithm} is changed to calculate Eq.~\ref{viterbi:multidet:vitsum}, the log-probability of a track at the geocentre ending in bin $j,k$ given that signal is in the real detector positions of $j,k+l$ and $j,k+m$. Line 14 is then modified so that $A_{j,k}$ stores the jump values, $i$, and the real detector positions, $l$ and $m$, which returned the highest probability.

% finding the most probable track
\item[Identification] The most probable track is identified in the same way as for the single detector case, first by finding the maximum value in the final time step of $V_{j,k}$ (line 19 in Alg.~\ref{viterbi:single:algorithm}). The track at the geocentre can then be found by iteratively following the jump values stored in $A_{j,k}$ back from this position. The track in each of the real detectors is determined by using the values of $l$ and $m$ indices also stored in $A_{j,k}$ to find the relative position of the track in each real detector compared to the geocentre.
\end{description}

This method can be extended to more than two detectors by including additional datasets and expanding the corresponding number dimensions of the maximisation procedures in the iterative steps.

%%%%%%%%%%%%%%%%%%%%%%%%%%%%%%%%%%%%%%%%%%%%%%%%%%%%%%%%%%%%%%%%%%%%%%%%%
%%%%%%%%%%%%%%%%%%%%%%%%%%%%%%%%%%%%%%%%%%%%%%%%%%%%%%%%%%%%%%%%%%%%%%%%%
\subsection{\label{viterbi:memory} Memory}
%%%%%%%%%%%%%%%%%%%%%%%%%%%%%%%%%%%%%%%%%%%%%%%%%%%%%%%%%%%%%%%%%%%%%%%%%
%
% general idea of memory
%
In this section we extend the basic Viterbi algorithm to improve its sensitivity to non-stochastic signals where there is some knowledge of its frequency evolution.
We do this by including a form of `memory' and this extension applies to both the single and multiple-detector cases.
Rather than considering only the previous step in our decision-making process, we now include the previous $m+1$ steps and expand the transition matrix to include these values.
A memory of $m=0$ therefore corresponds to the methods described in previous sections.
With a non-zero memory the transition matrix can a-priori make certain sequences of jumps more probable and assign different prior probabilities for these jump sequences e.g., `up then centre' may be less preferable to `centre then centre'.
As a result we can increase the chance of the most probable track matching an expected astrophysical signal.
In a single detector search with a memory of $m=1$, if we only allow \ac{UCD} transitions, then for every frequency bin we save 3 values. These are proportional to the log-probabilities of a track coming from a \ac{UCD} bin in the previous time step, where the maximisation is over the corresponding \ac{UCD} bins two time steps back.
Eq.~\ref{viterbi:multidet:vitsum} then is then modified to,
\begin{equation}
\label{viterbi:memory:stat}
V_{j,k,s} = \max_{h} ({C}_{j,k} + T_{s,h} +  V_{j-1,k+s,k+s+h}),
\end{equation}
where $s$ and $h$ refer to the \ac{UCD} jumps at the time step $j-1$ and $j-2$ respectively.  Similar to the previous two sections, the algorithm is split into three parts: initialisation, iteration, and the track identification:

\begin{description}
% initialisation
\item [Initialisation] The initialisation process needs to populate the first $m+1$ steps before the main iteration can start. At the first time step, the elements $V_{0,k,s}$ are set to the log-likelihoods $C_{0,k}$ as in Sec.~\ref{viterbi:single}.  There is no previous time step, so the element $s$ is not relevant. At the second time step, $V_{1,k,s}$ is calculated using Eq.~\ref{viterbi:memory:stat}, where there is no maximisation over $h$, it is assumed to be $0$, or a center jump. As there is no data before $j=0$, the maximisation at this point will always return the jump which has the largest prior probability, which in this case is a center jump. Therefore, the maximisation returns the same value for all frequency bins and can be set to a center jump.

%Iteration
\item [Iteration] For all following time steps the values for each element of $V_{j,k,s}$ in Eq.~\ref{viterbi:memory:stat} are calculated. This quantity is proportional to the log-probability of the track ending in time-frequency bin $j,k$, which was in the previous position of $j-1,k+s$. The corresponding value of $h$ that maximised the log-probability of the track is recorded in $A_{j,k,s}$.

%Identification
\item [Identification] The most probable track is identified in a similar way to the non-memory cases, by finding the highest-valued last element, $V_{N-1,k,s}$. The values of $s$ and $h$ are then followed back to find the most probable track. As an example, let us assume the most probable track finishes in bin $j,k,s = 10,5,0$, where the value of $m$ is $A_{10,5,0} = 1 = \rm{up}$. The previous position is then $j,k,s=10-1,5+s,m =10-1,5+0,1=9,5,1$ with a value $A_{9,5,1} = 0 = \text{Center}$, and the next track position is $j,k,s=9-1,5+1,0=8,6,0$ etc. The values of $j,k$ along this track describes most probable path.
\end{description}

The number of elements over which one must search increases rapidly with memory length, and has a strong impact on the computational cost of the analysis. For the single detector Viterbi approach the number of calculations made is $3 \times N \times M$ if we only allow \ac{UCD} jumps, where $N$ and $M$ are the number of time are frequency bins respectively. When memory is included this increases to $3^{m+1} \times N \times M $.

%%%%%%%%%%%%%%%%%%%%%%%%%%%%%%%%%%%%%%%%%%%%%%%%%%%%%%%%%%%%%%%%%%%%%%%%%
%%%%%%%%%%%%%%%%%%%%%%%%%%%%%%%%%%%%%%%%%%%%%%%%%%%%%%%%%%%%%%%%%%%%%%%%%
\subsection{\label{viterbi:sumdata}Summed input data}
%%%%%%%%%%%%%%%%%%%%%%%%%%%%%%%%%%%%%%%%%%%%%%%%%%%%%%%
%
% describe what we propose to do with summed SFT power
%
In this section we outline a method of incoherently-summing a set of \acp{SFT} to increase the \ac{SNR} of a signal in a segment. To be more precise, we actually sum the log-likelihoods, i.e. the quantity in Eq.~\ref{viterbi:periodogram}. We can write the new summed set of data $F_j$ as,
\begin{equation}
F_j = \sum_{i}^{N_s}C_{i,k}
\end{equation}
where $N_s$ is the number of \acp{SFT} to sum together and the log-likelihood $C(\nu_{i,k})$ is defined in Eq.~\ref{viterbi:periodogram}.
We can see this is possible by looking at Eq.~\ref{viterbi:single:likelihood}, where we can use the product of likelihoods,
\begin{equation}
\begin{split}
p(D \mid \nu) &\propto p(x_1,x_2 \ldots x_n \mid \nu) \\
&\propto p(x_1 \mid \nu) \ldots p(x_n \mid \nu) \\
&\propto \exp{\left( \sum_i C_{j,k}\right)}.
\end{split}
\end{equation}
If the data contains gaps where the detector was not observing, then we fill the gaps in the power spectrum with a constant value which is the expectation value of the log-likelihood. The procedure of filling in the gaps of the data is completed before any summing.  Therefore, the data should have the same mean regardless of how much real data is in each sum. In the examples that follow, we sum the \acp{SFT} over the length of one day.

The main motivation for summing the data is to increase the \ac{SNR} of a signal in the segments. The risk is that a signal can move between adjacent frequency bins during a day. To reduce this risk, we choose the frequency bin width such that it is more likely that a signal will be contained within a single frequency bin than cross a bin edge. In practise, to ensure that this is true, the segment or \ac{SFT} length and the number of segments which are summed can be tuned for each search. As well as increasing the \ac{SNR}, summing over one day should average out the antenna pattern. This means that the log-likelihood value in any bin should be more similar between detectors, however, there is still some variation due to the sky localisation and polarisation.

This also has two main effects on the transition matrix, the first is that as each segment of data is now one day long, a jump between frequency bins is far more likely, therefore, the transition matrix elements are modified to account for this. The second is that as the data is averaged over one day, the signal should remain is the same frequency bin between detectors, therefore, there is no longer a need for the multi-dimensional transition matrix described in Sec.~\ref{viterbi:multidet}.

The volume of the data is also reduced by a factor of $1/N_s$, therefore, the time taken for the algorithm to run is also reduced by the same factor.

%%%%%%%

%%%%%%%%%%%%%%%%%%%%%%%%%%%%%%%%%%%%%%%%%%%%%%%%%%%%%%%%%%%%%%%%%%%%%%%%%
%%%%%%%%%%%%%%%%%%%%%%%%%%%%%%%%%%%%%%%%%%%%%%%%%%%%%%%%%%%%%%%%%%%%%%%%%
\subsection{\label{viterbi:las}Line-aware statistic}
%%%%%%%%%%%%%%%%%%%%%%%%%%%%%%%%%%%%%%%%%%%%%%%%%%%%%%%%%%%%%%%%%%%%%%%%%
%
% what is the single detector line aware statistic
%
The single-detector algorithm described in Sec.~\ref{viterbi:single} returns the most probable track of the loudest signal assumed to be in Gaussian noise. However, an astrophysical signal is not expected to have an amplitude which is orders of magnitude above the noise floor, but have an amplitude more similar to the noise. Therefore, a signal with a large amplitude is more likely to be of instrumental origin rather than astrophysical \cite{Coughlin2010,Aasi:2015mqd,PhysRevD.97.082002}.

We first consider the model of Gaussian noise with no signal present. Within
a single summed segment, the likelihood of Gaussian noise at
frequency $\nu$ is given by a $\chi^2$ distribution,
\begin{equation}
\label{las:central}
p(F_j|\nu_j,M_{\text{N}},I) = \frac{1}{2^{d/2}\Gamma(d/2)}F_j^{d/2 - 1}\exp{\left\{
\frac{F_j}{2}\right\}}
\end{equation}
where $F_j$ is the frequency domain power summed over sub-segments within a single day, as described in Sec.~\ref{viterbi:sumdata} and  $d$ is the number of degrees of freedom,  equal to twice the total number of summed SFTs.  $M_{\rm{N}}$ represents the model that the data is simply Gaussian noise. In the presence of a signal (model $M_{\text{S}}$), the power should follow a non central $ \chi^2 $ distribution in which the non-centrality parameter $\lambda$ is the square of the \ac{SNR}, $(\lambda = \rho_{\rm{opt}}^2 )$, i.e.
\begin{equation}
\label{las:noncentral}
\begin{split}
p(F_j|\nu_j,\lambda,M_{\text{S}},I) = \frac{1}{2} \exp{\left\{ -\frac{F_j+\lambda}{2}\right\}} \left( \frac{F_j}{\lambda} \right)^{d/4 - 1/2} \\
I_{d/2 -1}\left( \sqrt{\lambda F_j}\right).
\end{split}
\end{equation}

If a signal is present we therefore expect the \ac{SFT} powers in the detector to follow Eq.~\ref{las:noncentral}.  We can then determine the evidence for model $M_{\text{S}}$ by marginalising over $\lambda$,
\begin{equation}
\label{las:signal:single}
\begin{split}
p(F^{(1)}_{j} \mid \nu_j,M_{\rm{S}},I) = \int_0^{\infty}  p(\lambda,w) 
p(F^{(1)}_{j}|\nu_j,\lambda,M_{\text{S}},I) d\lambda.
\end{split}
\end{equation}
Here we set the prior on $\lambda$ to be an exponential distribution of width $w$, this is done somewhat arbitrarily as we expect the majority of signals to have a low \ac{SNR}. This distribution follows,
\begin{equation}
\label{las:prior}
p(\lambda,w) = \exp\left( \frac{-\lambda}{w}\right).
\end{equation}

In this single-detector case, we expect an astrophysical signal to look very similar to that of a line other than its amplitude (or SNR). Therefore, we set the evidence for an astrophysical signal and an instrumental signal to follow Eq.~\ref{las:signal:single}, where the width $w$ different between the two models.

We then have three models, one for an astrophysical signal, one for an instrumental line and one for Gaussian noise. 

The posterior probability of model $M_{\text{GL}}$, which contains the probability of Gaussian noise or Gaussian noise with a line (taken as mutually exclusive) is
\begin{equation}
\begin{split}
p(M_{\rm{GL}} \mid F^{(1)}_{j},\nu_j ,I) = p(M_{\rm{G}} \mid F^{(1)}_{j},\nu_j ,I) \\
+p(M_{\rm{L}} \mid F^{(1)}_{j} ,\nu_j, I).
\end{split}
\end{equation}

We can now find the posterior odds ratio for the presence of a signal over noise or a line,
\begin{widetext}
\begin{equation}
\label{viterbi:odds:single}
\begin{split}
O^{(1)}_{\rm{S/GL}}(F^{(1)}_{j}\mid\nu_j) &=  \frac{p(M_{\rm{S}} \mid F^{(1)}_{j} ,\nu_j)}{p(M_{\rm{GL}} \mid F^{(1)}_{j},\nu_j)}
= \frac{p(M_{\rm{S}} \mid F^{(1)}_{j} ,\nu_j)}{p(M_{\rm{G}} \mid F^{(1)}_{j} ,\nu_j) +p(M_{\rm{L}} \mid F^{(1)}_{j} ,\nu_j)}\\
&=\frac{p(M_{\rm{S}})p(F^{(1)}_{j} \mid M_{\rm{S}},\nu_j)}{p(M_{\rm{G}})p(F^{(1)}_{j}\mid M_{\rm{G}},\nu_j) + p(M_{\rm{L}})p(F^{(1)}_{j}\mid M_{\rm{L}},\nu_j) } \\
&= \frac{p(F^{(1)}_{j} \mid M_{\rm{S}},\nu_j)p(M_{\rm{S}})/p(M_{\rm{G}})}{p(F^{(1)}_{j}\mid M_{\rm{G}},\nu_j) + p(F^{(1)}_{j}\mid M_{\rm{L}},\nu_j)p(M_{\rm{L}})/p(M_{\rm{G}}) }
\end{split}
\end{equation}
\end{widetext}
In practice it is convenient to use the log odds ratio,
\begin{equation}
\begin{split}
\label{viterbi:logodds:single}
\log\left[ O^{(1)}_{\rm{S/GL}}(F^{(1)}_{j})\right] &=  \log\left[ p(F^{(1)}_{j} \mid M_{\rm{S}}) \right] \\
&- \left[ \log\left( p(F^{(1)}_{j}\mid M_{\rm{G}}) \right. \right. \\
&\left.\left.+  p(F^{(1)}_{j}\mid M_{\rm{L}})p(M_{\rm{L}})/p(M_{\rm{G}})\right) \right]
\end{split}
\end{equation}
As we are only interested in the maximum of $\log\left[ O^{(1)}_{\rm{S/GL}}(F^{(1)}_{j})\right]$, the factor $\log\left[ p(M_{\rm{S}})/p(M_{\rm{G}})\right]$ can be dropped from the expression.

In this version of the Viterbi algorithm, rather than storing a value proportional to the log-probabilities as in Sec.~\ref{viterbi:multidet}, here we store a value proportional to the log-odds ratio.
Here we take the log-odds ratio defined in Eq.~\ref{viterbi:logodds:single} and add the log-prior odds $p(\bm{\nu} \mid M_S)/(p(\bm{\nu} \mid M_N) + p(\bm{\nu} \mid M_L))$ which is the log-prior or any particular track. By assuming that the track transitions for the line and noise model are equally probable for any jump, we set the denominator of the prior-odds is a constant $b$.
This then means Eq.~\ref{viterbi:single:vitsum} is modified to,
\begin{equation}
\begin{split}
\label{lineaware:stat}
\hat{V}_{i,j} = \max_{k,l,m}\left(T_{k,l,m} + b + V_{i-1,j+k}   \right. \\
 + \left.  \log\left[O^{(1)}_{\rm{S/GL}}\left(F^{(1)}_{j}\right)\right]\right),
\end{split}
\end{equation}
where $\hat{V}$ refers to a log-odds ratio.
The maximised statistic now has three tuneable parameters: the width ($w_S$) in Eq.~\ref{las:prior}, on the prior for a signal \ac{SNR} squared, $p_{\rm S}(\lambda)$, the width ($w_L$) of the prior in the case of a line, $p_{\rm L}(\lambda)$, and the ratio of the prior on the line and noise models, $p({M_{\rm L}})/p({M_{\rm G}})$.  These parameters are optimised for each search, where we initially estimate the \ac{SNR} of a signal we hope to be sensitive to in each time slice, then use this as a guide for the width of the signal prior. This is then repeated for an expected line \ac{SNR} and this is used for the width of the line prior. The ratio of line and noise models runs in the range 0 to 1, we set this limit as we do not expect an instrumental line to be as likely as Gaussian noise in any particular frequency bin.

\

%
% Two detector line aware statistic
%
This line-aware statistic can be applied in a more powerful way when we use multiple detectors and is similar to the approach in \cite{PhysRevD.89.064023}. The multiple-detector algorithm described in Sec.~\ref{viterbi:multidet} returns the most probable track of a common signal assumed to be in Gaussian noise. As a consequence the algorithm will return large values of the log-likelihood even if there are inconsistent values of \ac{SFT} power between the detectors, either from non-Gaussian noise or because the signal is not equally strong in the two detectors. However a signal with unequal power in the two detectors is more likely to be a non-Gaussian instrumental line than an astrophysical signal. The line-aware statistic described in this section is designed to make the search more robust to such instrumental artefacts within realistic non-Gaussian data whilst maintaining sensitivity to astrophysical signals.

%
% More detail on what this is applied to
%
For most of the analysis examples presented here we use data which is the incoherent sum of 30-minute normalised \acp{SFT} over a day (described in more detail in Sec.~\ref{viterbi:sumdata}). As a result the effects of the detector antenna patterns and of differential Doppler shifts are significantly reduced, and any signal should have a broadly similar summed log-likelihood in the same frequency bin in each detector. The statistic can then be modified such that we expect a similar log-likelihood in each detector.

In a similar way to the single-detector case, we can write out the evidence for each of the three models as follows. If a signal is present we therefore expect the \ac{SFT} powers in both detectors to follow Eq.~\ref{las:noncentral}.  Assuming for the moment that the noise variance is the same in both, we can determine the evidence for model $M_{\text{S}}$ by marginalising over $\lambda$,
\begin{equation}
\label{las:signal}
\begin{split}
p(F^{(1)}_{j},F^{(2)}_{j} \mid \nu_j,M_{\rm{S}},I) = \int_0^{\infty}  p(\lambda,w_{\rm_S}) \\
p(F^{(1)}_{j}|\nu_j,\lambda,M_{\text{S}},I)p(F^{(2)}_{j}|\nu_j,\lambda,M_{\text{S}},I) d\lambda.
\end{split}
\end{equation}
We set the prior on $\lambda$ the same as in the single detector case in Eq.~\ref{las:prior}.
In this case, if an instrumental line is present in one of the detectors we expect to see signal-like power in that detector and noise-like power in the other.  The evidence for this `line' model ($M_{\text{L}}$) is therefore
\begin{equation}
\label{las:line}
\begin{split}
p(F^{(1)}_{j},F^{(2)}_{j} \mid \nu_j,M_{\rm{L}},I) = \int_0^{\infty}  p(\lambda,w_{\rm_L}) \\
\left[ p(F^{(1)}_{j}|\nu_j,M_{\rm{N}},I)p(F^{(2)}_{j}|\nu_j,\lambda,M_{\rm{S}},I) \right. \\
\left. + p(F^{(1)}_{j}|\nu_j,\lambda,M_{\rm{S}},I)p(F^{(2)}_{j}|\nu_j,M_{\rm{N}},I)\right]d\lambda ,
\end{split}
\end{equation}
The third option is the simple case of approximately Gaussian noise in both of the detectors,
\begin{equation}
\label{las:noise}
\begin{split}
p(F^{(1)}_{j},F^{(2)}_{j} \mid \nu_j,\lambda,M_{\rm{G}},I) = p(F^{(1)}_{j} \mid \nu_j,M_{\rm{G}},I) \\
p(F^{(2)}_{j} \mid \nu_j,M_{\rm{G}},I) .
\end{split}
\end{equation}

We can now find the posterior odds ratio for the presence of a signal over noise or a line by following the same steps as in Eq.~\ref{viterbi:odds:single}. Once again we write this as a log-odds ratio,
\begin{equation}
\begin{split}
\label{viterbi:logodds}
\log\left[ O^{(2)}_{\rm{S/GL}}(F^{(1)}_{j},F^{(2)}_{j})\right] &=  \log\left[ p(F^{(1)}_{j},F^{(2)}_{j} \mid M_{\rm{S}}) \right] \\
&- \left[ \log\left( p(F^{(1)}_{j},F^{(2)}_{j}\mid M_{\rm{G}}) \right. \right. \\
&\left.\left.+  p(F^{(1)}_{j},F^{(2)}_{j}\mid M_{\rm{L}})p(M_{\rm{L}})/p(M_{\rm{G}})\right) \right]
\end{split}
\end{equation}
The factor $\log\left[ p(M_{\rm{S}})/p(M_{\rm{G}})\right]$ can again be dropped from the expression.

For the multi-detector case we then modify Eq.~\ref{viterbi:multidet:vitsum} to,
\begin{equation}
\begin{split}
\label{lineaware:stat:multi}
\hat{V}_{i,j} = \max_{k,l,m}\left(T_{k,l,m} + b + V_{i-1,j+k}   \right. \\
 + \left.  \log\left[O^{(2)}_{\rm{S/GL}}\left(F^{(1)}_{j},F^{(2)}_{j}\right)\right]\right),
\end{split}
\end{equation}
where $\hat{V}$ refers to a log-odds ratio.
This is then optimised over the same three parameters as the single detector case.

%%%%%%%%%%%%%%%%%%%%%%%%%%%%%%%%%%%%%%%%%%%%%%%%%%%
%%%%%%%%%%%%%%%%%%%%%%%%%%%%%%%%%%%%%%%%%%%%%%%%%%%%
\section{\label{results} Testing the algorithm}
\begin{figure*}
%\centering
\includegraphics[scale=0.5]{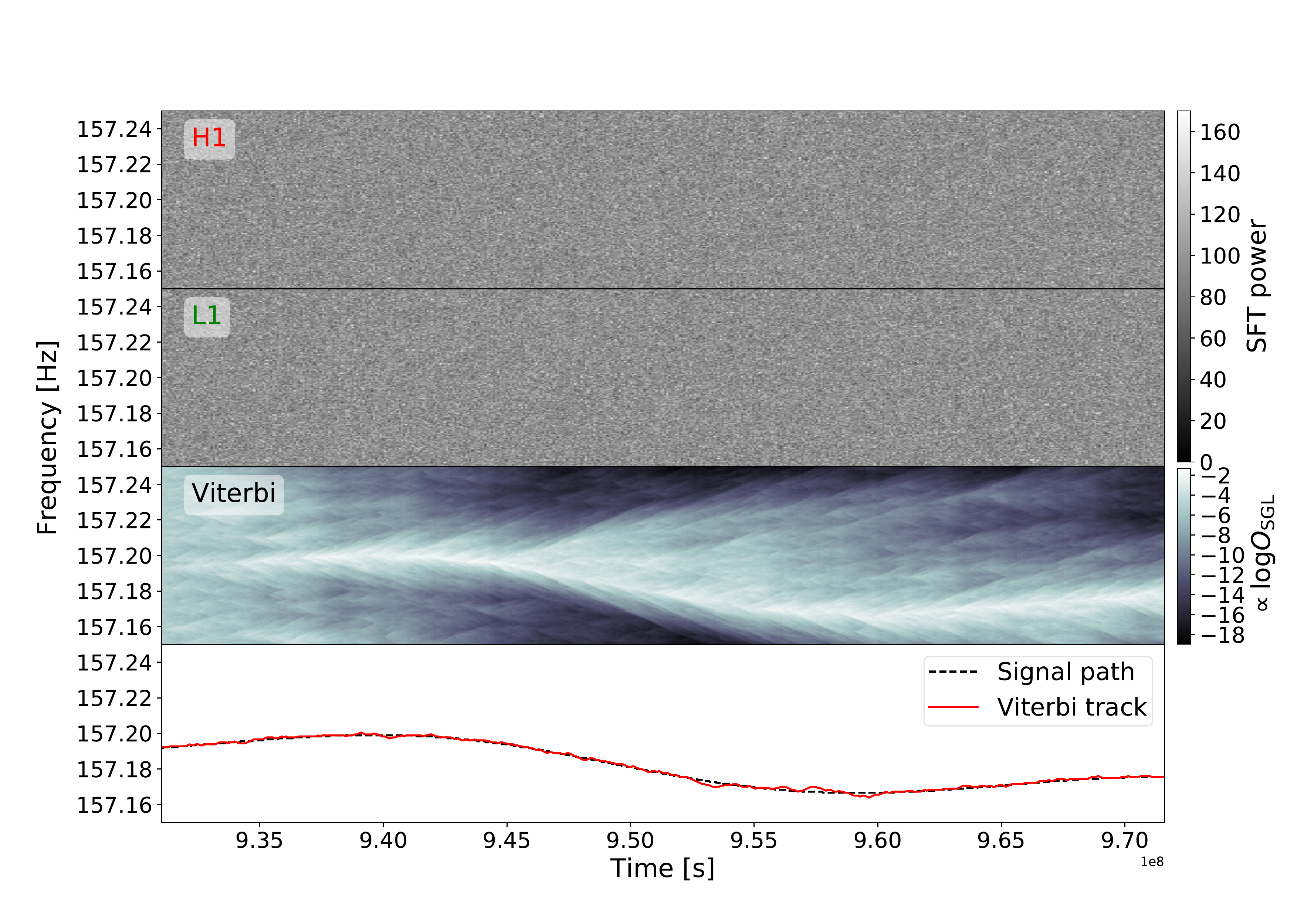}
\caption{\label{viterbi:tracks} The results that the SOAP algorithm returns from an injection with an optimal
\ac{SNR} of 90, i.e., the \ac{SNR} in H1 is 64 and the \ac{SNR} in L1 is 62.
The signal is injected into Gaussian noise, where the 1800\,s \acp{SFT} have been
summed over 1 day.  The top panel shows a simulation of summed \acp{SFT} from H1, the second panel shows the same for L1,
the third panel shows the values proportional to the
log-odds ratios in Eq.~\ref{lineaware:stat}.
The log-odds have been normalised such that the sum of
all the odds ratios in every time bin are equal to 1. The bottom panel shows the injected signal track (black dotted) and the track found in the `imaginary' detector by the two-detector SOAP search with the line-aware statistic (red), both of these tracks are at the geo-centre. In this case the \ac{RMS} of the difference between the Viterbi track and injected signal track was $\sim$1 bin.}
\end{figure*}

%%%%%%%%%%%%%%%%%%%%%%%%%%%%%%%%%%%%%%%%%%%%%%%%%%%%%
%
% Introduction to what we are testing on
%
The sensitivity of the algorithm was tested by searching for artificial
signals from isolated pulsars added to three types of noise-like data:
continuous Gaussian noise, Gaussian noise but with periods of missing data,
and real detector data (the S6 \ac{MDC}~\cite{Walsh2016}). The S6 \ac{MDC} refers to a standardised set of simulated signals which are injected into real data, this set is also what is used for the injections into the two Gaussian noise cases. We describe each
of the tests in more detail in Sec.~\ref{gaplessgauss},\ref{gausss6} and
\ref{results:s6}, but several common pre-processing steps are performed
before running these datasets through the Viterbi algorithm:
%
% Steps of the viterbi search
%
\begin{enumerate}
\item We read \acp{SFT} generated from 1800\,s stretches of data in 2\,Hz
    bands between 100 and 200\,Hz. The \acp{SFT} length is chosen to ensure that any signal is likely to be contained within the width of a single
    frequency bin during the length of one day, rather than being split
    across the bin edges (see below).
\item We estimate the noise \ac{PSD} for each \ac{SFT} by calculating a
    running median over frequency using LALSuite
    code {\tt XLALSFTtoRnmed} \cite{lalsuite}, this includes a bias factor to
    convert this to the mean and has a width of 100 bins. We then normalise the \ac{SFT} by dividing it
    by its running median, giving the noise-like parts of the spectrum a
    mean power of approximately one.

\item The \acp{SFT} are then summed over one day, as described in
    Sec.~\ref{viterbi:sumdata}. The signal parameters are chosen so that
    within the frequencies of the search, the signal will not fall in more than two frequency bins over this
    period.

    The differential Doppler shift of a signal seen at two detector sites due to the Earth's rotation $\Delta f^{(1,2)}_{\rm{rot}}$ is simply
\begin{equation} \label{results:doppler}
\Delta f^{(1,2)}_{\rm{rot}} = \frac{({{\bm v}^{(1)}} -{{\bm v}^{(2)}})\cdot
\hat{\bm s}}{c} f_0,
\end{equation}
where ${\bm v}^{(1,2)}$ is the velocity of detector $1,2$ in an inertial reference frame, $f_0$ is the
instantaneous signal frequency in the frame, $\hat{\bm s}$ is the unit vector in the direction of the source and  $c$ is the speed of light.
The maximum difference in frequency seen by the two \ac{LIGO} detectors
is 
\begin{equation} \label{results:doppler:diff}
\Delta f_{\rm rot} \approx 6.5\times 10^{-7} f_0,
\end{equation}
so the frequency measured from a source in the equatorial plane with $f_0=200$\,Hz will differ by up to $1.3 \times 10^{-4}$\,Hz in the two detectors.
This is $\sim 4$ times smaller than the frequency bin width of 1800\,s \acp{SFT} ($5.6 \times
10^{-4}$\,Hz), so signals at frequencies lower than this are likely to appear in the same frequency bin in the two detectors. Therefore, whilst at higher frequencies we still allow the signal to be in different frequency bins between the detectors, in the following searches, we do not allow this.

\item The data is then split into 0.1\,Hz sub-bands which are overlapping by 0.05\,Hz. These were chosen to ensure that signals are contained within a sub-band over the year. On these timescales the important contributions to the frequency evolution are the spin-down rate of the pulsar and the Doppler shift due to the earth orbit.
To investigate the doppler shift, we can look at a signal at 200\,Hz, using Eq.~\ref{results:doppler} we can calculate the maximum shift in frequency due to the earths orbit as,
\begin{equation}
\Delta f_{\rm orbit} = \frac{2 \pi R_o}{T_o} \frac{1}{c} f_0 \approx 9.9 \times 10^{-5} f_0,
\end{equation}
where $T_{\text{o}}$ and $R_{\text{o}}$ are the
earth orbit time and radius. This gives a maximum doppler shift of $0.019 \; \rm{Hz}$, this is a $\sim 1/5$ of the width of a sub-band, therefore, is more likely to be totally contained within a sub-band than crossing over the edge.
To account for the cases where the signal frequency crosses over the edge of a sub-band, the sub-bands overlap by 0.05\,Hz so that the majority of the signals should be completely contained within at least one of the sub-bands.
To investigate the spin-down of the pulsar, we look at the length of data, $T=4.05 \times 10^7$\,s and we choose a sub-band width of 0.1\,Hz. For a signal to drift over the width of a whole sub-band we would need f-dot of,
\begin{equation}
\frac{df}{dt} > \left|\frac{-0.1}{4.05 \times 10^7}\right| = 2.4 \times 10^{-9} \rm{Hz/s}.
\end{equation}
The majority of the injections that follow satisfy this condition, signals which are greater than this, and therefore drift over multiple bands, are vetoed from the search.
\item The two detector Viterbi algorithm is then run using the line aware
statistic (see Sec.~\ref{viterbi:las}). There are 4 parameters which we optimise in this search. The transition probabilities, where we have one parameter $\tau$ which is the ratio of the probability
of going straight to the probability of going either up or down. Due to the averaging
procedure, the signals received at each detector are forced to follow a common track which is equal to the `imaginary' detectors track. The other three parameters, $w_S, w_L \; \rm{and} \; p(M_L)/p(M_N)$, are described in Sec.~\ref{viterbi:las}.
\item The algorithm then returns the most probable track though the data, and the value
$\propto$ the log-odds in the final time step, i.e., the
maximum final value, $\max_j(V_{N,j})$, in Eq.~\ref{lineaware:stat}, which is then our detection statistic.

\end{enumerate}
%
% Example plot of what viterbi gives
%
As an example of what the algorithm returns, Fig.~\ref{viterbi:tracks} shows
the tracks in the two detectors, H1 and L1. This also shows the
log-odds ratio of ending in any
frequency bin, i.e., all the elements in Eq.~\ref{lineaware:stat}.  In this figure, each time segment of the odds ratios have been normalised such that the sum of the odds ratios is 1.

In the following tests there are two main quantities which we use to determine
the sensitivity. These are sensitivity depth $\mathcal{D}$ and the optimal
\ac{SNR} $\rho$.  The sensitivity depth, $\mathcal{D}$, is defined in
\cite{Behnke2015} as,
\begin{equation}
\label{sigmoid}
\mathcal{D}(f) = \frac{\sqrt{S_h(f)}}{h_0},
\end{equation}
where $S_h(f)$ is the single-sided noise \ac{PSD} and $h_0$ is the \ac{GW} amplitude.  The optimal \ac{SNR} is defined as,
\begin{equation}
\rho^2 = \sum_X 4
\Re\int^{\infty}_{\infty}\frac{\tilde{h}^X(f)\tilde{h}^{X*}(f)}{S^X(f)}df,
\end{equation}
 where $X$ indexes the
detectors and $\tilde{h}(f)$ is the Fourier transform of the time
series of the signal $h(t)$. This expression is defined in~\cite{Prix2007} for
a double-sided \ac{PSD} and we have defined it for the more common single-sided
case.

%%%%%%%%%%%%%%%%%%%%%%%%%%%%%%%%%%%%%%%
%%%%%%%%%%%%%%%%%%%%%%%%%%%%%%%%%%%%%
\subsection{\label{gaplessgauss}S6 injections into gapless Gaussian noise}
%%%%%%%%%%%%%%%%%%%%%%%%%%%%%%%%%%%%%%%
%%%%%%%%%%%%%%%%%%%%%%%%%%%%%%%%%%%%%%
%
% intro to what data was used and defenitions
%
The first test involves injecting signals into Gaussian noise. The power spectrum of a Gaussian noise time-series follows a $\chi^2$ distribution with two degrees of freedom, therefore, as we search through the power spectrum, we generate spectrograms which follow a $\chi^2$ distribution. 
These spectrograms are 0.1\,Hz wide and are set at 0.05\,Hz intervals between 100\,Hz and 200\,Hz. The bins are $1./1800$\,Hz wide and 1800s long, where the total length of data is the same as S6, i.e. $\sim 1.3$ years.
We then generate the signals, where the pulsars parameters are
fixed to the same values as the injections in the S6 \ac{MDC} in this band, 
these values are outlined in \cite{Walsh2016}.

%
% describe the pre-pruning of signals
%
The values of $f_0$ for the injections were not always centred in a
sub-band, therefore a number of sub-bands contained only part of the injected
signal. These sub-bands were ignored as they contaminated the signal statistics
and only the sub-band which contained the whole signal was accepted.  This
reduced the number of sub-bands from 2000 to 1762 with the removal of 238
sub-bands containing only part of a signal.
This set also includes signals that
drift across multiple sub-bands due to their high spin-down
rate. Only two signals were removed due to their spin-down values, which were $> 5 \times 10^{-9}$\,Hz/s, these were the two hardware injections in the 100-200\,Hz band.

%
% how roc curves and efficiencies are generated/found
%
For each injection the SOAP algorithm returns the detection statistic described in Sec.~\ref{viterbi:las} and \ref{results}.
We calculate a false alarm rate, which is the fraction of bands that have no injection that do exceed a given threshold. This is set to 1\% and is used as a detection threshold.
We then take all of the bands and if they pass the threshold we set them as detected, i.e. 1, and if they do not they are set as not detected, i.e. 0.
This then leaves us with a set of binomial data, where the efficiency curves later in the paper are sigmoids which have been fitted to this.
The sigmoid follows,
\begin{equation}
s(x; x_0, k)  = \frac{1}{1-\exp{(-k(x - x_0))}}.
\end{equation}
The fit is done by sampling the posterior, i.e.,
\begin{equation}
\label{binomialbayes}
p(x_0, k \mid b) \propto  p(x_0,k)p(x \mid x_0, k),
\end{equation}
where $p(x_0,k)$ is the prior and we set to a flat prior and $p(x \mid x_0, k)$ is the likelihood function which is defined by,
\begin{equation}
\begin{split}
p(\bar{x} \mid x_0, k) = \prod_{j=0}\frac{n!}{k!(n-k)!}s(x_j \mid x_0, k)^{k} \\ (1-s(x_j \mid x_0,k))^{n-k}.
\end{split}
\end{equation}
To plot the efficiency curves and lower and upper error bounds, we sample Eq.~\ref{binomialbayes} using \ac{MCMC} and then take the mean and the 5th and 95th percentiles respectively for each point in SNR or depth and plot these.
Fig.~\ref{gausss6:eff_snr} and  \ref{gausss6:eff_depth} then shows the efficiency curves for the analyses plotted against the signals optimal \ac{SNR} and depth respectively.
The parameters of the search and their optimised values are shown in Tab.~\ref{gaplessgauss:pars}. Where we set the prior on the line model to 0 as this part is irrelevant to this search due to the lack of lines in the data.

\begin{table}
\centering
\caption{Table shows the ranges of the search parameters for the gapless Gaussian injections search Sec.~\ref{gaplessgauss} and the optimised parameters. There are 10 parameter values spaced linearly between the limits. \label{gaplessgauss:pars}}

\bgroup
\def\arraystretch{1.5}
\centering
\begin{tabular}{c c c c c}
\hline
\hline
 & $\tau$ & $w_S$ & $w_L$ & $p(M_L)/p(M_N)$ \\
\hline
limits & [1.0,1.3]& [0.1,5.0]& None& 0.0\\
\hline
optimised & 1.1 & 2.06 & None & 0.0 \\
\hline
\end{tabular}
\egroup
\end{table}
From this we can determine that in Gaussian noise without gaps, the Viterbi
algorithm can detect to an \ac{SNR} of $\sim60 $ and a depth of $\sim
33$\,Hz$^{-1/2}$ with 95\% efficiency at a 1\% false alarm.

Fig.~\ref{gausss6:res_snr} and \ref{gausss6:res_depth},
show the \ac{RMS} of the difference between the injected signal track and the track found by Viterbi for \ac{SNR} and sensitivity depth respectively. This shows that at \ac{SNR} of 60, where we are detecting signals with a 95\% efficiency, the signals have a mean \ac{RMS} of $\sim 2$ frequency bins. \

%%%%%%%%%%%%%%%%%%%%%%%%%%%%%%%%%%%%%
%%%%%%%%%%%%%%%%%%%%%%%%%%%%%%%%%%%%%
\subsection{\label{gausss6} S6 injections into Gaussian noise with gaps}
%%%%%%%%%%%%%%%%%%%%%%%%%%%%%%%%%%%
%%%%%%%%%%%%%%%%%%%%%%%%%%%%%%%%%%%

%
% Intro to what data was used
%
In the second test, we attempt to more closely mirror the S6 \ac{MDC}~\cite{Walsh2016} in two stages. The first uses the same injection method as Sec.~\ref{gaplessgauss} however, removes the \acp{SFT} where there are gaps in S6. 
The second uses the same injection method again including gaps, however, uses a different value for the noise floor for each \ac{SFT}, this is calculated for each band and \ac{SFT} from S6 data.

Both detectors in S6 had a duty cycle of $\sim$50\%~\cite{DetCharS6}, which means that there are sections of time where there is no data in either one or both detectors. In the sections where one detector is observing but the other is not, the multi detector statistic will not behave correctly as it only has access to data from a single detector.
In these sections we switch from using the multi-detector statistic to the single-detector statistic using the same parameters, these are both defined in defined in Sec.~\ref{viterbi:las}.

%
% describe the pre-pruning of signals and how efficieny curves were calculated
%
The process of removing sub-bands and generating efficiency curves is the same as in Sec.~\ref{gaplessgauss}.

\begin{figure*}
\centering

\begin{subfigure}[h]{\columnwidth}
\centering
\includegraphics[scale=0.33]{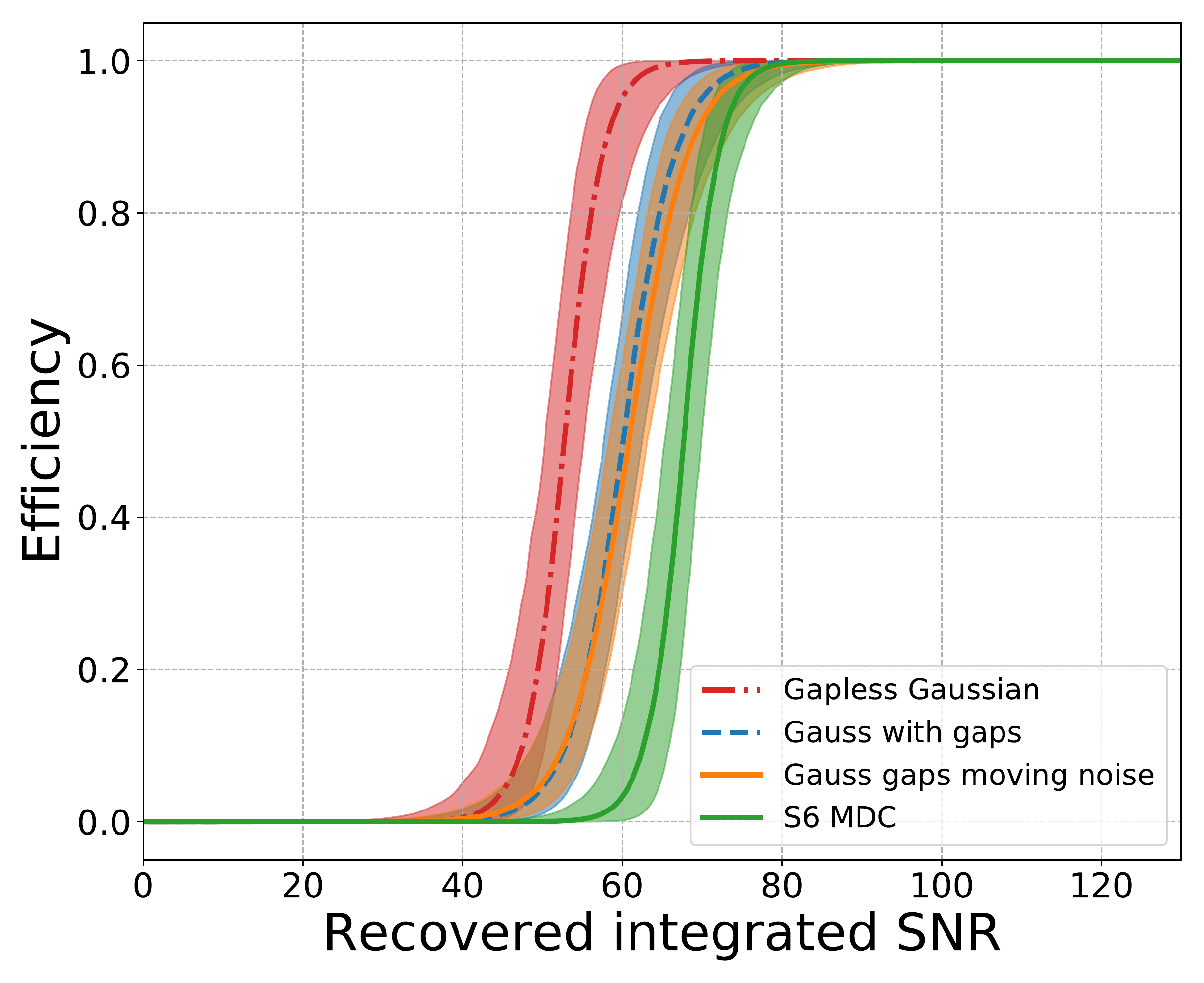}
\subcaption{}
\label{gausss6:eff_snr}
\end{subfigure}
\begin{subfigure}[h]{\columnwidth}
\includegraphics[scale=0.33]{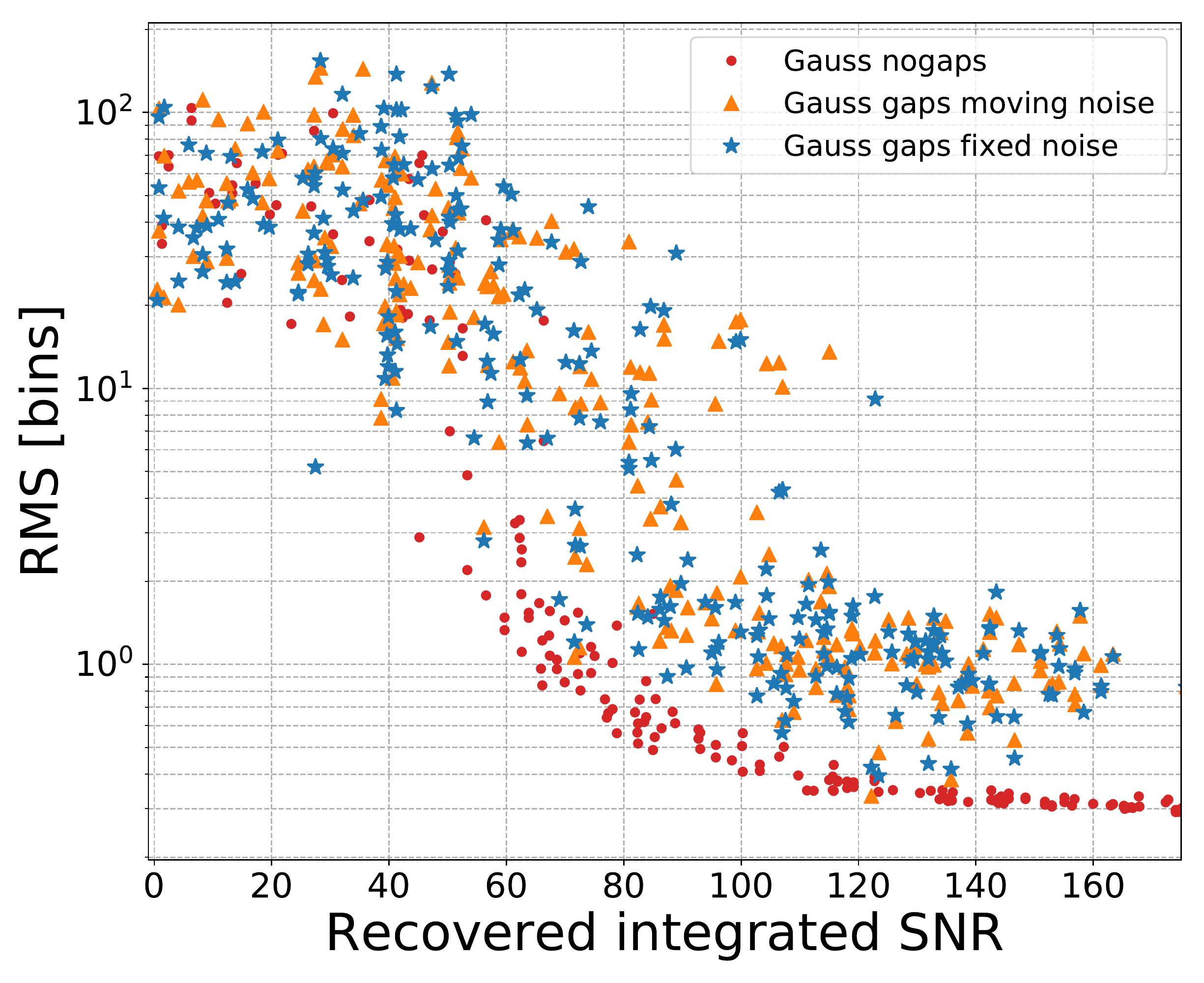}
\subcaption{}
\label{gausss6:res_snr}
\end{subfigure}

\begin{subfigure}[h]{\columnwidth}
\includegraphics[scale=0.33]{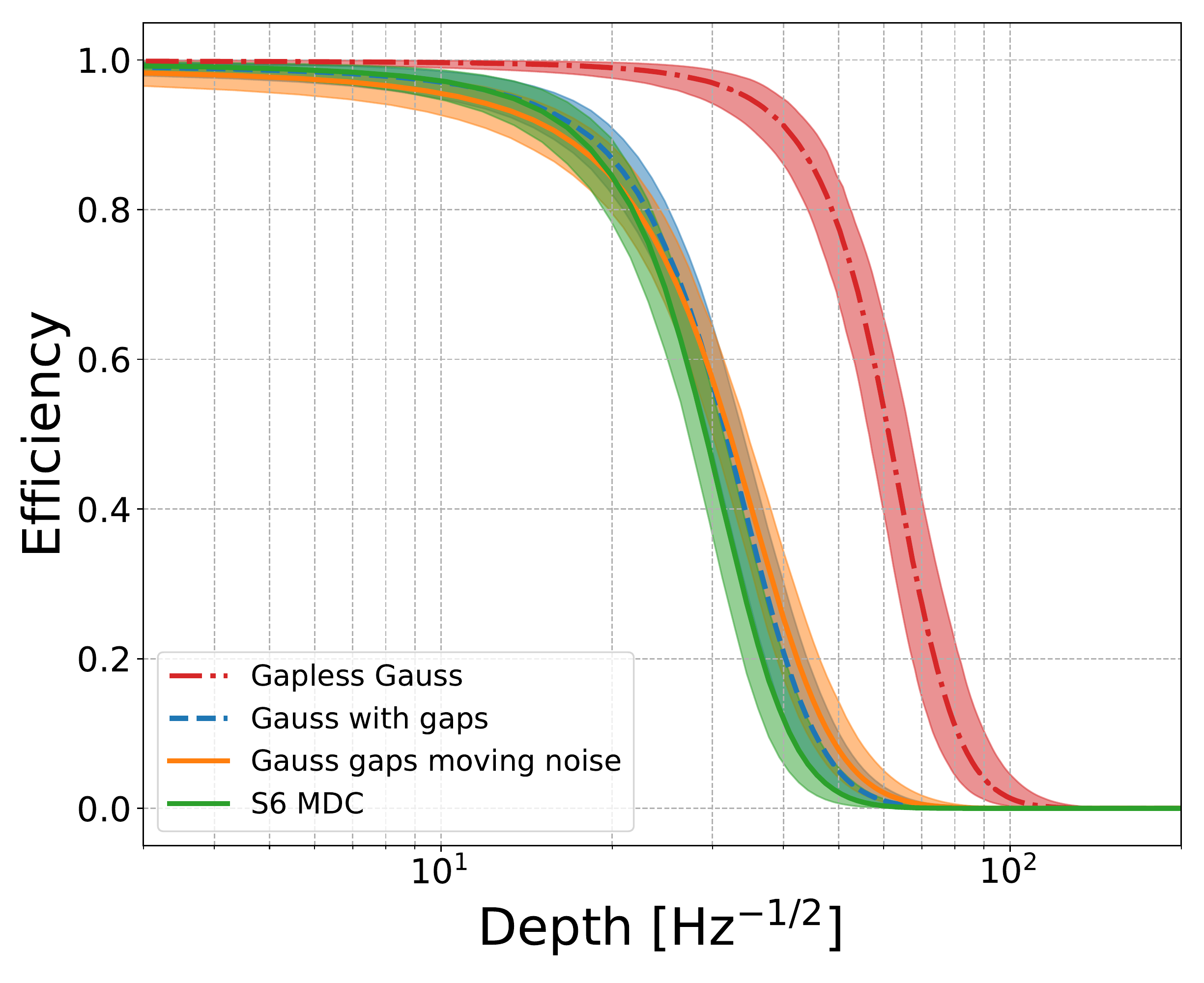}
\subcaption{}
\label{gausss6:eff_depth}
\end{subfigure}
\begin{subfigure}[h]{\columnwidth}
\includegraphics[scale=0.33]{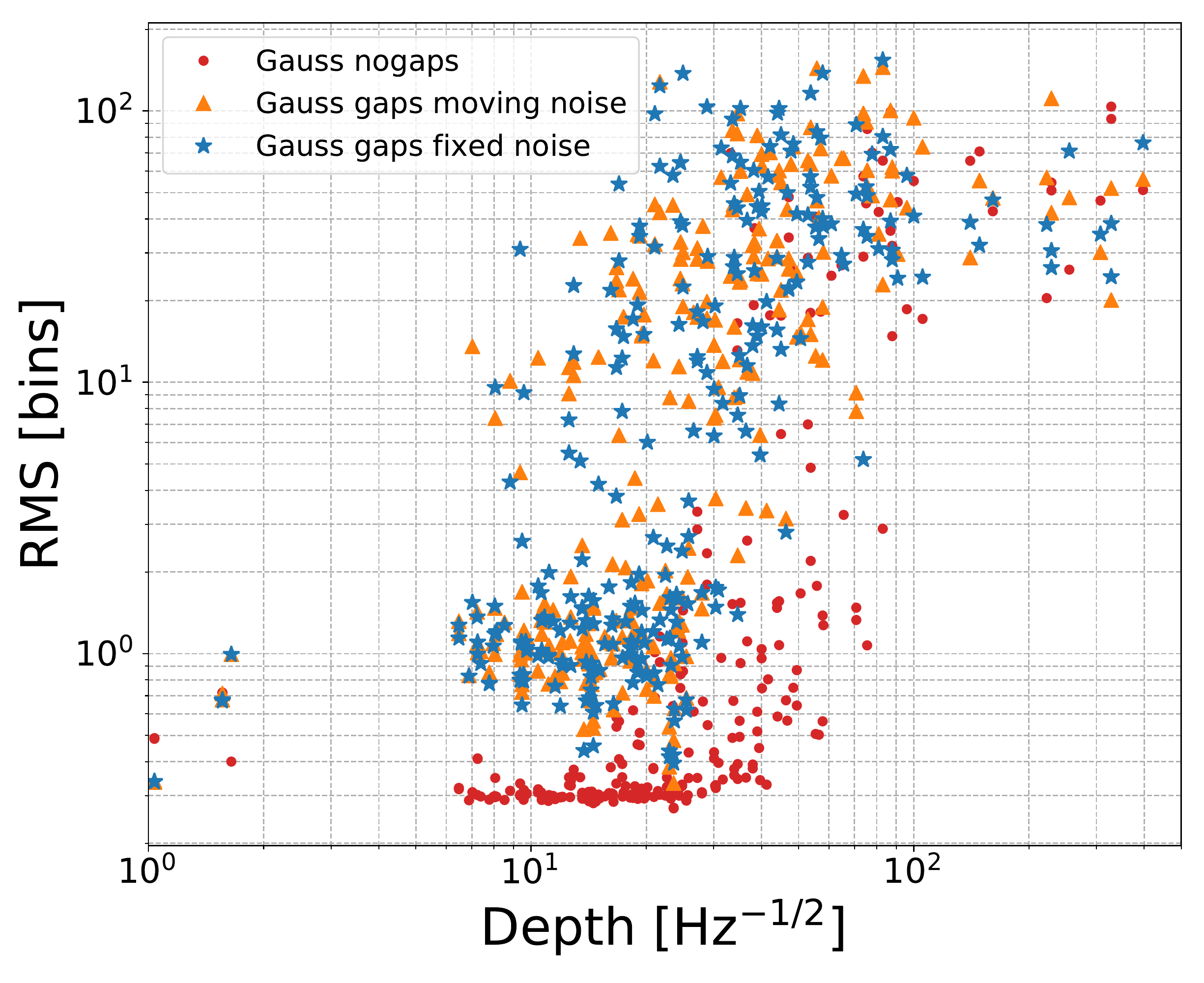}
\subcaption{}
\label{gausss6:res_depth}
\end{subfigure}

\caption{Panels \ref{gausss6:eff_snr} and \ref{gausss6:eff_depth} show the detection efficiency as a function of \ac{SNR} and depth respectively. Here \ac{SNR} is the the integrated \ac{SNR} which we would expect to recover from the available data. The four curves refer to injections into gapless Gaussian noise (red), Gaussian noise with gaps in data, where the noise floor is either fixed (blue-dashed) or it is moving with time (orange) in the same way as the S6~\ac{MDC} and injections into real data i.e. the S6~\ac{MDC}.
The curves are made by fitting a sigmoid Eq.~\ref{sigmoid} to binomial detection data with a 1\% false alarm rate, as explained in Sec.~\ref{gaplessgauss}, the error bounds are the 5\% and 95\% intervals.
At 95\% efficiency and a 1\% false alarm rate, this shows we can detect to an \ac{SNR}  of $\sim 60$ and a sensitivity depth of $\sim 34$\,Hz$^{-1/2}$ for gapless Gaussian noise and an \ac{SNR}  of $\sim 69$ and $72$ and a sensitivity depth of $\sim 13$\,Hz$^{-1/2}$ and $\sim 10$\,Hz$^{-1/2}$ for the Gaussian with gaps case with fixed noise floor and moving noise floor respectively. For the S6 \ac{MDC} we can detect an \ac{SNR} of $\sim 74$ and a sensitivity depth of $\sim 13$\,Hz$^{-1/2}$. 
 Panels \ref{gausss6:res_snr} and \ref{gausss6:res_depth} show the \ac{RMS} of the difference between the injected signal track and the track found by SOAP as a function of \ac{SNR} and sensitivity depth respectively. \label{gausss6:plots} }
\end{figure*}

We set a 1\% false alarm rate and generate an efficiency curve for
\ac{SNR} and depth in Fig.~\ref{gausss6:eff_snr} and
Fig.~\ref{gausss6:eff_depth} respectively. From these efficiency plots we can
see to an \ac{SNR} of $\sim 72$ or a depth of $\sim 13$\,Hz$^{-1/2}$ at a 95\%
confidence with a false alarm of 1\%.

The parameters of the search which were optimised and their optimised values are shown in Tab.~\ref{gausss6:pars}.

\begin{table}
\centering
\caption{Table shows the ranges of the search parameters for the S6-like injections into Gaussian noise with gaps Sec.~\ref{results:s6} and the optimised parameters. The parameters were spaced linearly between the limits. \label{gausss6:pars} }

\bgroup
\def\arraystretch{1.5}
\centering
\begin{tabular}{c c c c c}
\hline
\hline
 & $\tau$ & $w_S$ & $w_L$ & $p(M_L)/p(M_N)$ \\
\hline
limits & [1.0,1.4]& [0.1,5.0]& None& 0.0\\
\hline
optimised & 1.1 & 2.06 & None & 0.0 \\
\hline
\end{tabular}
\egroup
\end{table}

In Fig.~\ref{gausss6:res_snr} and ~\ref{gausss6:res_depth} show the \ac{RMS} of the difference between the injected signal track and the track found by Viterbi for \ac{SNR} and sensitivity depth respectively. This shows that at \ac{SNR} of 72, where we are detecting signals with a 95\% efficiency, the signals have a mean \ac{RMS} of $\sim 10$ frequency bins.

%%%%%%%%%%%%%%%%%%%%%%%%%%%%%%%%%%%%%%%%%%%%%%%%%%%%%%%%%%%%%%%%%%%%%%%%%%%%%%%%
\subsection{\label{results:s6}Tests on the S6 MDC}
%%%%%%%%%%%%%%%%%%%%%%%%%%%%%%%%%%%%%%%%%%%%%%%%%%%%%%%%%%%%%%%%%%%%%%%%%%%%%%%%
%
% Intro to what data was run on and how it was split
%
For a more direct comparison to other \ac{CW} searches and to see how the
algorithm performs with real data, we test the two detector SOAP algorithm using the S6 \ac{MDC}. We focus this search on the 100-200\,Hz band, there are two main reasons for this, one being that this is \acp{LIGO} most sensitive band and the other is that for much higher frequencies the signal will drift over larger frequency ranges, therefore, our \ac{SFT} length will have to be changed.
Here the 1800\,s \acp{SFT} are split as in Sec.~\ref{results}, whereafter normalisation, the data is split into 0.1\,Hz wide sub-bands overlapping by 0.05\,Hz.

The two detector SOAP algorithm using the line-aware statistic in Sec.~\ref{viterbi:las} is then run on each sub-band under the assumption that the detectors have the same sensitivity.
For this search we have four parameters which we optimise, the ranges and optimised values are shown in Tab.~\ref{s6:pars}.

\begin{table}
\centering
\caption{Table shows the ranges of the search parameters for the S6 search Sec.~\ref{results:s6} and the optimised parameters. The parameters, $\tau$, $w_{\rm{L}}$ and $w_{\rm{S}}$ were distributed in log space between the limits and $p(M_L)/p(M_N)$ is distributed uniformly. \label{s6:pars} }

\bgroup
\def\arraystretch{1.5}
\centering
\begin{tabular}{c c c c c}
\hline
\hline
 & $\tau$ & $w_S$ & $w_L$ & $p(M_L)/p(M_N)$ \\
\hline
limits & [1.0,1.1]& [0.1,5.0]& [0.1,6.0]& [0.0,1.0]\\
\hline
optimised & 1.00000001 & 4.0 & 5.0 & 0.0387 \\
\hline
\end{tabular}
\egroup
\end{table}
%

%
% Process of vetoing lines
%
As in Sec.~\ref{gausss6}, only the sub-bands which contained the entire frequency evolution of the signal were selected.
Out of the 2000 sub-bands, 238 were removed due the sub-band only containing part of the signals frequency evolution.
The main difference between the analysis for Gaussian noise and real data is that
the real data is contaminated with instrumental lines. This means that whilst the techniques described in Sec.~\ref{viterbi:las} reduce the number of contaminated bands with a high statistic value, there are still
instrumental lines which are coincident between the detectors and which could not be removed
with these techniques. Within the data there are large number of lines at integer Hertz, which are seen in coincidence between the two detectors, these are thought to originate from digital electronics \cite{Coughlin2010}. Therefore the frequency bins $\pm1$ bin of each integer frequency in
Hertz were removed and filled with the expectation value of the noise. To remove instrumental effects at other frequencies,
the sub-bands which gave values of our
statistic above a chosen threshold were investigated by eye. In this case 344 sub-bands were
investigated, and any which were contaminated were vetoed. 
From these 344 sub-bands, 193 were removed from the analysis. The predominant feature in the bands which were removed were broad spectral features which lasted the whole run. Therefore, out of the 2000 sub-bands which are searched over, a total number of 431 sub-bands were removed.

%
% generating efficiency curves
%
The process to calculate the efficiency curves is the same as
in Sec.~\ref{gausss6} and \ref{gaplessgauss}.

%
% Describe the results plots
%

Fig.~\ref{gausss6:eff_depth} and Fig.~\ref{gausss6:eff_snr} show the efficiency curves
for \ac{SNR} and depth respectively.  These show that we can detect and
\ac{SNR} of $\sim 74$ and a sensitivity depth of $\sim 13$\,Hz$^{-1/2}$ with
an efficiency of 95\% at a false alarm of 1\%. These results can then be compared to other searches in the S6 MDC comparison paper~\cite{Walsh2016}. Whilst we only search in the 100 - 200 Hz range, the closest comparison in \cite{Walsh2016} is the test in the 40 - 500 Hz range, such as in Fig.~4 in \cite{Walsh2016}. Here our algorithm sits roughly in the middle of all other searches in terms of sensitivity.

%%%%%%%%%%%%%%%%%%%%%%%%%%%%%%%%%%%%%%%%%%%%%%%%%%%%%%%%%%%%%%%%%%%%%%%%%%%%%%%
%%%%%%%%%%%%%%%%%%%%%%%%%%%%%%%%%%%%%%%%%%%%%%%%%%%%%%%%%%%%%%%%%%%%%%%%%%%%%%%
\subsection{\label{results:time}Computational cost}
%%%%%%%%%%%%%%%%%%%%%%%%%%%%%%%%%%%%%%%%%%%%%%%%%%%%%%%%%%%%%%%%%%%%%%%%%%%%%%%
%%%%%%%%%%%%%%%%%%%%%%%%%%%%%%%%%%%%%%%%%%%%%%%%%%%%%%%%%%%%%%%%%%%%%%%%%%%%%%%

One of the main strengths of this search is the drastically reduced
computational cost when compared to other current \ac{CW} searches.
The scaling of the computing cost can be estimated for a single detector by looking at the number of calculations that need to be made. The number of calculations for a single detector search, $N^{(1)}_{\rm{calcs}}$ is,
\begin{equation}
\label{results:time:singlecalc}
N^{(1)}_{\rm{calcs}} = n_{1}^{m}NM,
\end{equation}
where $n_1$ is the size of the transition matrix, $N$ is the number of \acp{SFT}, $M$ is the number of frequency bins and $m$ is the amount of memory described in Sec.~\ref{viterbi:memory}. Where the computing cost scales linearly with the number of frequency bins and \acp{SFT}.
In the following test we ignore `memory' and look at the time taken for the single detector search where the time taken to read and save data is ignored. Here the data is the same size as the S6 \ac{MDC} for a single detector search and the search is over a 0.1\,Hz band, where we set $n_1=3$. This test, and the following test, was run locally on a MacBook Air with a 1.3 GHz Intel Core i5 processor .We can then write the time taken ,$T$, as,
\begin{equation}
T = 0.56\,\rm{sec}\left( \frac{N}{22538}\right) \left( \frac{M}{180}\right) \left( \frac{N_{\rm{bands}}}{1}\right),
\end{equation}

where  $N_{\rm{bands}}$ is the number of different frequency
bands.
For the multiple, $Q$, detector case, we can then generalise Eq.~\ref{results:time:singlecalc} and write the number of calculations $N^{(Q)}_{\rm{calcs}}$ as,
\begin{equation}
\label{results:time:numbercalcs}
N^{(Q)}_{\rm{calcs}} = NMn_{1}^{m} \prod_{q=1}^{Q}n_{q+1},
\end{equation}
where $n_1$ is the first dimension of the transition matrix, $Q$ is the number of detectors and $n_{q+1}$ is the size of the transition matrix element which refers to detector $q$.
For our tests we set $n_1=n_{q+1}=3$ and use 2 detectors i.e. $Q=2$ which each have the same size data as the previous test. The actual time taken to run however, depends on the version of the algorithm which is run. For example, including the line aware statistic slows the search slightly.
For the two detector where two \ac{SFT} powers are summed,
\begin{equation}
T = 1.35 \rm{s}\left( \frac{N}{22538}\right) \left( \frac{M}{180}\right) \left( \frac{N_{\rm{bands}}}{1}\right).
\end{equation}
The same search now including the line aware statistic, which is implemented using a lookup table, changes this to,
\begin{equation}
T = 25.7 \rm{s}\left( \frac{N}{22538}\right) \left( \frac{M}{180}\right) \left( \frac{N_{\rm{bands}}}{1}\right).
\end{equation}

%%%%%%%%%%%%%%%%%%%%%%%%%%%%%%%%%%%%%%%%%%%%%%%%%%%%%%%%%%%%%%%%%%%%%%%%%%%%%%%
%%%%%%%%%%%%%%%%%%%%%%%%%%%%%%%%%%%%%%%%%%%%%%%%%%%%%%%%%%%%%%%%%%%%%%%%%%%%%%%
\section{\label{dicussion}Discussion}
%%%%%%%%%%%%%%%%%%%%%%%%%%%%%%%%%%%%%%%%%%%%%%%%%%%%%%%%%%%%%%%%%%%%%%%%%%%%%%%
%%%%%%%%%%%%%%%%%%%%%%%%%%%%%%%%%%%%%%%%%%%%%%%%%%%%%%%%%%%%%%%%%%%%%%%%%%%%%%%

In this paper we describe an application of the Viterbi algorithm, called SOAP, to
search for continuous sources of gravitational waves.
This paper outlines the method and derives the statistics behind the method in a consistent Bayesian formalism. It then presents the results from the first set of tests of sensitivity for the SOAP algorithm on three separate datasets.

%
% overview of results from tests
%
We tested SOAP on a set of fake isolated pulsar signals in the
100\;--\;200\,Hz range, based on 1800s \acp{SFT} summed over one day.
The three datasets that inclded these signals comprised continuous Gaussian noise, Gaussian noise but with temporal gaps corresponding to LIGO dead times in the S6 data run, and real data, i.e. the
S6 \ac{MDC}. Although a major attraction of SOAP is its sensitivity to a wide
range of signal types, in the tests above it was optimised to detect isolated pulsar signals below 100\,Hz with low spin-down to offer a comparison with other \ac{CW} searches. From these tests, by setting a
95\% efficiency and a false alarm of 1\%, we found that in the case of  continuous Gaussian data we could detect a signal with an optimal \ac{SNR} of $\sim 60$ and a
depth of $\sim 33$\,Hz$^{-1/2}$ with an \ac{RMS} of the difference between the injected and Viterbi track being $\sim 2$ frequency bins.
When gaps were introduced into the data to simulate S6 we could detect a signal with an
\ac{SNR} $\sim 72$  and a depth of $\sim 10$\,Hz$^{-1/2}$, with an \ac{RMS} of $\sim 10$ bins.  The drop in sensitivity here is simply because  there is $\sim 50 \%$ less data compared to the previous case. Finally, in the S6 \ac{MDC} we could
detect a signal with an \ac{SNR} $\sim 74$ and a depth of $\sim
13$\,Hz$^{-1/2}$.
These real data contain non-Gaussian artefacts such as instrumental lines and this causes a further drop in sensitivity.
Whilst not a full comparison to other searches in the S6 \ac{MDC} \cite{Walsh2016}, as we only tested on a subset
of the bands, this search has a sensitivity which is comparable to some other \ac{CW} searches, however offers a massive increase in speed.

We chose the specific frequency band to search over as the data which we used, i.e. the summed data, becomes less effective at frequencies much higher than 200\,Hz, and using the parameters of our simulations, signals can spread over many frequency bins in a day, reducing sensitivity further, however this can be mitigated by using shorter \acp{SFT} or performing their summation over 12 (rather than 24) hours.

%
% some other features to test in the future
%
The methods described in this paper present a basic approach for gravitational-wave signal searches using SOAP. However there are several further developments that could increase its sensitivity. Some of these are outlined below:

One of the main features which reduces the sensitivity of the search is non-gaussianities within the data, namely instrumental lines. Although we have a statistic which penalises these features, in some cases it will also penalise a strong signal. For example, when the amplitude of the noise floor is high for one detector or the duty factor is lower, the signal will appear more like an instrumental line to this statistic. We hope to improve the search statistic in the future by searching for consistent amplitudes as opposed to consistent \ac{SNR}, i.e the statistic will take the amplitude of the noise floor and the duty factor into account.

One variation of this method which has been described in this paper is `memory', which is where the tracks jump in frequency is determined by the previous $n$ jumps. This has yet to be fully tested, however, we expect that this will increase our sensitivity to signals where have a better idea of their frequency evolution. This however, comes at a cost in computational time which we can estimate given Eq.~\ref{results:time:numbercalcs} in Sec.~\ref{results:time}.

Further additions to the search include using the Fourier transform of the \ac{SFT} power along the Viterbi track as a detection statistic.
If the Viterbi track follows that from an astrophysical signal, then we should see the effects of the antenna pattern in this Fourier transform as a peak at half a sidereal day.
If the track follows something which is not astrophysical then this should not be seen this peak in this Fourier transform.
This only applies to the search directly on the \acp{SFT} not the summed data, as the antenna pattern variations will have been averaged out in the summing.

As well as searching for astrophysical signals, SOAP can also be used to search for and identify instrumental lines. Here we use single detector data, or multiple channels from a single detector, to identify quasi-monochromatic features on the data for further study.

Whilst this paper presents initial tests on sensitivity, further tests will be needed for a full comparison to other \ac{CW} search methods.  
This search, however, aims to look for signals which may not follow the standard
frequency evolution and is intended to return potentially interesting
candidates for a more sensitive followup.

\section{Acknowledgements}

We would like to acknowledge Dr Matt Pitkin for his help during this project, as well as the continuous wave working group of \ac{LIGO}-Virgo Collaboration. This research is supported by the Science and Technology Facilities Council., G.W. and C.M. are supported by the Science and Technology Research Council (grant No. ST/L000946/1) . The authors are grateful for computational resources provided by the LIGO Laboratory and Cardiff University and supported by National Science Foundation Grants PHY-0757058 and PHY-0823459 and STFC grant ST/I006285/1.

% not sure how to abbreviate author list here???
\bibliography{soap}

%merlin.mbs apsrev4-1.bst 2010-07-25 4.21a (PWD, AO, DPC) hacked
%Control: key (0)
%Control: author (8) initials jnrlst
%Control: editor formatted (1) identically to author
%Control: production of article title (-1) disabled
%Control: page (0) single
%Control: year (1) truncated
%Control: production of eprint (0) enabled
\begin{thebibliography}{32}%
\makeatletter
\providecommand \@ifxundefined [1]{%
 \@ifx{#1\undefined}
}%
\providecommand \@ifnum [1]{%
 \ifnum #1\expandafter \@firstoftwo
 \else \expandafter \@secondoftwo
 \fi
}%
\providecommand \@ifx [1]{%
 \ifx #1\expandafter \@firstoftwo
 \else \expandafter \@secondoftwo
 \fi
}%
\providecommand \natexlab [1]{#1}%
\providecommand \enquote  [1]{``#1''}%
\providecommand \bibnamefont  [1]{#1}%
\providecommand \bibfnamefont [1]{#1}%
\providecommand \citenamefont [1]{#1}%
\providecommand \href@noop [0]{\@secondoftwo}%
\providecommand \href [0]{\begingroup \@sanitize@url \@href}%
\providecommand \@href[1]{\@@startlink{#1}\@@href}%
\providecommand \@@href[1]{\endgroup#1\@@endlink}%
\providecommand \@sanitize@url [0]{\catcode `\\12\catcode `\$12\catcode
  `\&12\catcode `\#12\catcode `\^12\catcode `\_12\catcode `\%12\relax}%
\providecommand \@@startlink[1]{}%
\providecommand \@@endlink[0]{}%
\providecommand \url  [0]{\begingroup\@sanitize@url \@url }%
\providecommand \@url [1]{\endgroup\@href {#1}{\urlprefix }}%
\providecommand \urlprefix  [0]{URL }%
\providecommand \Eprint [0]{\href }%
\providecommand \doibase [0]{http://dx.doi.org/}%
\providecommand \selectlanguage [0]{\@gobble}%
\providecommand \bibinfo  [0]{\@secondoftwo}%
\providecommand \bibfield  [0]{\@secondoftwo}%
\providecommand \translation [1]{[#1]}%
\providecommand \BibitemOpen [0]{}%
\providecommand \bibitemStop [0]{}%
\providecommand \bibitemNoStop [0]{.\EOS\space}%
\providecommand \EOS [0]{\spacefactor3000\relax}%
\providecommand \BibitemShut  [1]{\csname bibitem#1\endcsname}%
\let\auto@bib@innerbib\@empty
%</preamble>
\bibitem [{\citenamefont {Abbott}\ \emph {et~al.}(2009)\citenamefont {Abbott}
  \emph {et~al.}}]{LIGO}%
  \BibitemOpen
  \bibfield  {author} {\bibinfo {author} {\bibfnamefont {B.~P.}\ \bibnamefont
  {Abbott}} \emph {et~al.},\ }\href
  {http://stacks.iop.org/0034-4885/72/i=7/a=076901} {\bibfield  {journal}
  {\bibinfo  {journal} {Reports on Progress in Physics}\ }\textbf {\bibinfo
  {volume} {72}},\ \bibinfo {pages} {076901} (\bibinfo {year}
  {2009})}\BibitemShut {NoStop}%
\bibitem [{\citenamefont {Aasi}\ \emph
  {et~al.}(2015{\natexlab{a}})\citenamefont {Aasi} \emph {et~al.}}]{LIGO2015}%
  \BibitemOpen
  \bibfield  {author} {\bibinfo {author} {\bibfnamefont {J.}~\bibnamefont
  {Aasi}} \emph {et~al.},\ }\href {\doibase 10.1088/0264-9381/32/7/074001}
  {\bibfield  {journal} {\bibinfo  {journal} {Classical and Quantum Gravity}\
  }\textbf {\bibinfo {volume} {32}},\ \bibinfo {pages} {074001} (\bibinfo
  {year} {2015}{\natexlab{a}})}\BibitemShut {NoStop}%
\bibitem [{\citenamefont {Acernese}\ \emph {et~al.}(2008)\citenamefont
  {Acernese} \emph {et~al.}}]{VIRGO}%
  \BibitemOpen
  \bibfield  {author} {\bibinfo {author} {\bibfnamefont {F.}~\bibnamefont
  {Acernese}} \emph {et~al.},\ }\href
  {http://stacks.iop.org/0264-9381/25/i=11/a=114045} {\bibfield  {journal}
  {\bibinfo  {journal} {Classical and Quantum Gravity}\ }\textbf {\bibinfo
  {volume} {25}},\ \bibinfo {pages} {114045} (\bibinfo {year}
  {2008})}\BibitemShut {NoStop}%
\bibitem [{\citenamefont {Acernese}\ \emph {et~al.}(2014)\citenamefont
  {Acernese} \emph {et~al.}}]{Acernese_2014}%
  \BibitemOpen
  \bibfield  {author} {\bibinfo {author} {\bibfnamefont {F.}~\bibnamefont
  {Acernese}} \emph {et~al.},\ }\href {\doibase 10.1088/0264-9381/32/2/024001}
  {\bibfield  {journal} {\bibinfo  {journal} {Classical and Quantum Gravity}\
  }\textbf {\bibinfo {volume} {32}},\ \bibinfo {pages} {024001} (\bibinfo
  {year} {2014})}\BibitemShut {NoStop}%
\bibitem [{\citenamefont {Prix}(2009)}]{Prix2009}%
  \BibitemOpen
  \bibfield  {author} {\bibinfo {author} {\bibfnamefont {R.}~\bibnamefont
  {Prix}},\ }\enquote {\bibinfo {title} {Gravitational waves from spinning
  neutron stars},}\ in\ \href {\doibase 10.1007/978-3-540-76965-1_24} {\emph
  {\bibinfo {booktitle} {Neutron Stars and Pulsars}}},\ \bibinfo {editor}
  {edited by\ \bibinfo {editor} {\bibfnamefont {W.}~\bibnamefont {Becker}}}\
  (\bibinfo  {publisher} {Springer Berlin Heidelberg},\ \bibinfo {address}
  {Berlin, Heidelberg},\ \bibinfo {year} {2009})\ pp.\ \bibinfo {pages}
  {651--685}\BibitemShut {NoStop}%
\bibitem [{\citenamefont {Owen}(2009)}]{Owen:2009fg}%
  \BibitemOpen
  \bibfield  {author} {\bibinfo {author} {\bibfnamefont {B.~J.}\ \bibnamefont
  {Owen}},\ }\href@noop {} {\  (\bibinfo {year} {2009})},\ \Eprint
  {http://arxiv.org/abs/0903.2603} {arXiv:0903.2603 [astro-ph.SR]} \BibitemShut
  {NoStop}%
%%CITATION = ARXIV:0903.2603;%%
\bibitem [{\citenamefont {Dupuis}\ and\ \citenamefont
  {Woan}(2005)}]{Dupuis2005}%
  \BibitemOpen
  \bibfield  {author} {\bibinfo {author} {\bibfnamefont {R.~J.}\ \bibnamefont
  {Dupuis}}\ and\ \bibinfo {author} {\bibfnamefont {G.}~\bibnamefont {Woan}},\
  }\href {\doibase 10.1103/PhysRevD.72.102002} {\bibfield  {journal} {\bibinfo
  {journal} {Physical Review D - Particles, Fields, Gravitation and Cosmology}\
  }\textbf {\bibinfo {volume} {72}} (\bibinfo {year} {2005}),\
  10.1103/PhysRevD.72.102002},\ \Eprint {http://arxiv.org/abs/0508096}
  {arXiv:0508096 [gr-qc]} \BibitemShut {NoStop}%
\bibitem [{\citenamefont {Astone}\ \emph {et~al.}(2010)\citenamefont {Astone},
  \citenamefont {D'Antonio}, \citenamefont {Frasca},\ and\ \citenamefont
  {Palomba}}]{Astone_2010}%
  \BibitemOpen
  \bibfield  {author} {\bibinfo {author} {\bibfnamefont {P.}~\bibnamefont
  {Astone}}, \bibinfo {author} {\bibfnamefont {S.}~\bibnamefont {D'Antonio}},
  \bibinfo {author} {\bibfnamefont {S.}~\bibnamefont {Frasca}}, \ and\ \bibinfo
  {author} {\bibfnamefont {C.}~\bibnamefont {Palomba}},\ }\href {\doibase
  10.1088/0264-9381/27/19/194016} {\bibfield  {journal} {\bibinfo  {journal}
  {Classical and Quantum Gravity}\ }\textbf {\bibinfo {volume} {27}},\ \bibinfo
  {pages} {194016} (\bibinfo {year} {2010})}\BibitemShut {NoStop}%
\bibitem [{\citenamefont {Jaranowski}\ \emph
  {et~al.}(1998{\natexlab{a}})\citenamefont {Jaranowski}, \citenamefont
  {Kr\'olak},\ and\ \citenamefont {Schutz}}]{PhysRevD.58.063001}%
  \BibitemOpen
  \bibfield  {author} {\bibinfo {author} {\bibfnamefont {P.}~\bibnamefont
  {Jaranowski}}, \bibinfo {author} {\bibfnamefont {A.}~\bibnamefont
  {Kr\'olak}}, \ and\ \bibinfo {author} {\bibfnamefont {B.~F.}\ \bibnamefont
  {Schutz}},\ }\href {\doibase 10.1103/PhysRevD.58.063001} {\bibfield
  {journal} {\bibinfo  {journal} {Phys. Rev. D}\ }\textbf {\bibinfo {volume}
  {58}},\ \bibinfo {pages} {063001} (\bibinfo {year}
  {1998}{\natexlab{a}})}\BibitemShut {NoStop}%
\bibitem [{\citenamefont {Collaboration}\ \emph {et~al.}(2017)\citenamefont
  {Collaboration}, \citenamefont {Collaboration}, \citenamefont {Buchner},
  \citenamefont {Cognard}, \citenamefont {Corongiu}, \citenamefont {Freire},
  \citenamefont {Guillemot}, \citenamefont {Hobbs}, \citenamefont {Kerr},
  \citenamefont {Lyne}, \citenamefont {Possenti}, \citenamefont {Ridolfi},
  \citenamefont {Shannon}, \citenamefont {Stappers},\ and\ \citenamefont
  {Weltevrede}}]{O1knownpulsar2017}%
  \BibitemOpen
  \bibfield  {author} {\bibinfo {author} {\bibfnamefont {L.~S.}\ \bibnamefont
  {Collaboration}}, \bibinfo {author} {\bibfnamefont {V.}~\bibnamefont
  {Collaboration}}, \bibinfo {author} {\bibfnamefont {S.}~\bibnamefont
  {Buchner}}, \bibinfo {author} {\bibfnamefont {I.}~\bibnamefont {Cognard}},
  \bibinfo {author} {\bibfnamefont {A.}~\bibnamefont {Corongiu}}, \bibinfo
  {author} {\bibfnamefont {P.~C.~C.}\ \bibnamefont {Freire}}, \bibinfo {author}
  {\bibfnamefont {L.}~\bibnamefont {Guillemot}}, \bibinfo {author}
  {\bibfnamefont {G.~B.}\ \bibnamefont {Hobbs}}, \bibinfo {author}
  {\bibfnamefont {M.}~\bibnamefont {Kerr}}, \bibinfo {author} {\bibfnamefont
  {A.~G.}\ \bibnamefont {Lyne}}, \bibinfo {author} {\bibfnamefont
  {A.}~\bibnamefont {Possenti}}, \bibinfo {author} {\bibfnamefont
  {A.}~\bibnamefont {Ridolfi}}, \bibinfo {author} {\bibfnamefont {R.~M.}\
  \bibnamefont {Shannon}}, \bibinfo {author} {\bibfnamefont {B.~W.}\
  \bibnamefont {Stappers}}, \ and\ \bibinfo {author} {\bibfnamefont
  {P.}~\bibnamefont {Weltevrede}},\ }\href
  {http://stacks.iop.org/0004-637X/839/i=1/a=12} {\bibfield  {journal}
  {\bibinfo  {journal} {The Astrophysical Journal}\ }\textbf {\bibinfo {volume}
  {839}},\ \bibinfo {pages} {12} (\bibinfo {year} {2017})}\BibitemShut
  {NoStop}%
\bibitem [{\citenamefont {Abbott}\ \emph
  {et~al.}(2019{\natexlab{a}})\citenamefont {Abbott} \emph
  {et~al.}}]{O2knownpulsar:2019}%
  \BibitemOpen
  \bibfield  {author} {\bibinfo {author} {\bibfnamefont {B.~P.}\ \bibnamefont
  {Abbott}} \emph {et~al.} (\bibinfo {collaboration} {LIGO Scientific,
  Virgo}),\ }\href@noop {} {\  (\bibinfo {year} {2019}{\natexlab{a}})},\
  \Eprint {http://arxiv.org/abs/1902.08507} {arXiv:1902.08507 [astro-ph.HE]}
  \BibitemShut {NoStop}%
%%CITATION = ARXIV:1902.08507;%%
\bibitem [{\citenamefont {Brady}\ and\ \citenamefont
  {Creighton}(2000)}]{PhysRevD.61.082001}%
  \BibitemOpen
  \bibfield  {author} {\bibinfo {author} {\bibfnamefont {P.~R.}\ \bibnamefont
  {Brady}}\ and\ \bibinfo {author} {\bibfnamefont {T.}~\bibnamefont
  {Creighton}},\ }\href {\doibase 10.1103/PhysRevD.61.082001} {\bibfield
  {journal} {\bibinfo  {journal} {Phys. Rev. D}\ }\textbf {\bibinfo {volume}
  {61}},\ \bibinfo {pages} {082001} (\bibinfo {year} {2000})}\BibitemShut
  {NoStop}%
\bibitem [{\citenamefont {Abbott}\ \emph
  {et~al.}(2019{\natexlab{b}})\citenamefont {Abbott} \emph
  {et~al.}}]{Pisarski:2019vxw}%
  \BibitemOpen
  \bibfield  {author} {\bibinfo {author} {\bibfnamefont {B.~P.}\ \bibnamefont
  {Abbott}} \emph {et~al.} (\bibinfo {collaboration} {LIGO Scientific,
  Virgo}),\ }\href@noop {} {\  (\bibinfo {year} {2019}{\natexlab{b}})},\
  \Eprint {http://arxiv.org/abs/1903.01901} {arXiv:1903.01901 [astro-ph.HE]}
  \BibitemShut {NoStop}%
%%CITATION = ARXIV:1903.01901;%%
\bibitem [{\citenamefont {Ellis}\ \emph {et~al.}(2006)\citenamefont {Ellis},
  \citenamefont {Heffernan}, \citenamefont {Gutierrez},\ and\ \citenamefont
  {Dalessandro}}]{soap}%
  \BibitemOpen
  \bibfield  {author} {\bibinfo {author} {\bibfnamefont {D.~R.}\ \bibnamefont
  {Ellis}}, \bibinfo {author} {\bibfnamefont {J.}~\bibnamefont {Heffernan}},
  \bibinfo {author} {\bibfnamefont {S.}~\bibnamefont {Gutierrez}}, \ and\
  \bibinfo {author} {\bibfnamefont {D.}~\bibnamefont {Dalessandro}},\ }\href
  {https://www.imdb.com/title/tt0417148/} {\enquote {\bibinfo {title} {Snakes
  on a plane},}\ } (\bibinfo {year} {2006})\BibitemShut {NoStop}%
\bibitem [{\citenamefont {Viterbi}(1967)}]{Viterbi1967}%
  \BibitemOpen
  \bibfield  {author} {\bibinfo {author} {\bibfnamefont {A.~J.}\ \bibnamefont
  {Viterbi}},\ }\href {\doibase 10.1109/TIT.1967.1054010} {\bibfield  {journal}
  {\bibinfo  {journal} {IEEE Transactions on Information Theory}\ }\textbf
  {\bibinfo {volume} {13}},\ \bibinfo {pages} {260} (\bibinfo {year}
  {1967})}\BibitemShut {NoStop}%
\bibitem [{\citenamefont {Suvorova}\ \emph {et~al.}(2016)\citenamefont
  {Suvorova}, \citenamefont {Sun}, \citenamefont {Melatos}, \citenamefont
  {Moran},\ and\ \citenamefont {Evans}}]{Suvorova2016}%
  \BibitemOpen
  \bibfield  {author} {\bibinfo {author} {\bibfnamefont {S.}~\bibnamefont
  {Suvorova}}, \bibinfo {author} {\bibfnamefont {L.}~\bibnamefont {Sun}},
  \bibinfo {author} {\bibfnamefont {A.}~\bibnamefont {Melatos}}, \bibinfo
  {author} {\bibfnamefont {W.}~\bibnamefont {Moran}}, \ and\ \bibinfo {author}
  {\bibfnamefont {R.~J.}\ \bibnamefont {Evans}},\ }\href {\doibase
  10.1103/PhysRevD.93.123009} {\bibfield  {journal} {\bibinfo  {journal}
  {Physical Review D - Particles, Fields, Gravitation and Cosmology}\ }\textbf
  {\bibinfo {volume} {93}},\ \bibinfo {pages} {1} (\bibinfo {year} {2016})},\
  \Eprint {http://arxiv.org/abs/1606.02412} {arXiv:1606.02412} \BibitemShut
  {NoStop}%
\bibitem [{\citenamefont {Sun}\ \emph {et~al.}(2018)\citenamefont {Sun},
  \citenamefont {Melatos}, \citenamefont {Suvorova}, \citenamefont {Moran},\
  and\ \citenamefont {Evans}}]{PhysRevD.97.043013}%
  \BibitemOpen
  \bibfield  {author} {\bibinfo {author} {\bibfnamefont {L.}~\bibnamefont
  {Sun}}, \bibinfo {author} {\bibfnamefont {A.}~\bibnamefont {Melatos}},
  \bibinfo {author} {\bibfnamefont {S.}~\bibnamefont {Suvorova}}, \bibinfo
  {author} {\bibfnamefont {W.}~\bibnamefont {Moran}}, \ and\ \bibinfo {author}
  {\bibfnamefont {R.~J.}\ \bibnamefont {Evans}},\ }\href {\doibase
  10.1103/PhysRevD.97.043013} {\bibfield  {journal} {\bibinfo  {journal} {Phys.
  Rev. D}\ }\textbf {\bibinfo {volume} {97}},\ \bibinfo {pages} {043013}
  (\bibinfo {year} {2018})}\BibitemShut {NoStop}%
\bibitem [{\citenamefont {Suvorova}\ \emph {et~al.}(2017)\citenamefont
  {Suvorova}, \citenamefont {Clearwater}, \citenamefont {Melatos},
  \citenamefont {Sun}, \citenamefont {Moran},\ and\ \citenamefont
  {Evans}}]{PhysRevD.96.102006}%
  \BibitemOpen
  \bibfield  {author} {\bibinfo {author} {\bibfnamefont {S.}~\bibnamefont
  {Suvorova}}, \bibinfo {author} {\bibfnamefont {P.}~\bibnamefont
  {Clearwater}}, \bibinfo {author} {\bibfnamefont {A.}~\bibnamefont {Melatos}},
  \bibinfo {author} {\bibfnamefont {L.}~\bibnamefont {Sun}}, \bibinfo {author}
  {\bibfnamefont {W.}~\bibnamefont {Moran}}, \ and\ \bibinfo {author}
  {\bibfnamefont {R.~J.}\ \bibnamefont {Evans}},\ }\href {\doibase
  10.1103/PhysRevD.96.102006} {\bibfield  {journal} {\bibinfo  {journal} {Phys.
  Rev. D}\ }\textbf {\bibinfo {volume} {96}},\ \bibinfo {pages} {102006}
  (\bibinfo {year} {2017})}\BibitemShut {NoStop}%
\bibitem [{\citenamefont {Abbott}\ \emph {et~al.}(2017)\citenamefont {Abbott}
  \emph {et~al.}}]{PhysRevD.95.122003}%
  \BibitemOpen
  \bibfield  {author} {\bibinfo {author} {\bibfnamefont {B.~P.}\ \bibnamefont
  {Abbott}} \emph {et~al.} (\bibinfo {collaboration} {LIGO Scientific
  Collaboration and Virgo Collaboration}),\ }\href {\doibase
  10.1103/PhysRevD.95.122003} {\bibfield  {journal} {\bibinfo  {journal} {Phys.
  Rev. D}\ }\textbf {\bibinfo {volume} {95}},\ \bibinfo {pages} {122003}
  (\bibinfo {year} {2017})}\BibitemShut {NoStop}%
\bibitem [{\citenamefont {Abbott}\ \emph {et~al.}(2018)\citenamefont {Abbott}
  \emph {et~al.}}]{Abbott:2018hgk}%
  \BibitemOpen
  \bibfield  {author} {\bibinfo {author} {\bibfnamefont {B.~P.}\ \bibnamefont
  {Abbott}} \emph {et~al.} (\bibinfo {collaboration} {LIGO Scientific,
  Virgo}),\ }\href@noop {} {\  (\bibinfo {year} {2018})},\ \Eprint
  {http://arxiv.org/abs/1810.02581} {arXiv:1810.02581 [gr-qc]} \BibitemShut
  {NoStop}%
%%CITATION = ARXIV:1810.02581;%%
\bibitem [{\citenamefont {{Sun}}\ and\ \citenamefont
  {{Melatos}}(2018)}]{2018arXiv181003577S}%
  \BibitemOpen
  \bibfield  {author} {\bibinfo {author} {\bibfnamefont {L.}~\bibnamefont
  {{Sun}}}\ and\ \bibinfo {author} {\bibfnamefont {A.}~\bibnamefont
  {{Melatos}}},\ }\href@noop {} {\bibfield  {journal} {\bibinfo  {journal}
  {arXiv e-prints}\ ,\ \bibinfo {eid} {arXiv:1810.03577}} (\bibinfo {year}
  {2018})},\ \Eprint {http://arxiv.org/abs/1810.03577} {arXiv:1810.03577
  [astro-ph.IM]} \BibitemShut {NoStop}%
\bibitem [{\citenamefont {Bretthorst}(1988)}]{Bretthorst1988}%
  \BibitemOpen
  \bibfield  {author} {\bibinfo {author} {\bibfnamefont {G.~L.}\ \bibnamefont
  {Bretthorst}},\ }\href@noop {} {\emph {\bibinfo {title} {Springer-Verlag}}}\
  (\bibinfo {year} {1988})\ p.\ \bibinfo {pages} {220}\BibitemShut {NoStop}%
\bibitem [{\citenamefont {Jaranowski}\ \emph
  {et~al.}(1998{\natexlab{b}})\citenamefont {Jaranowski}, \citenamefont
  {Krolak},\ and\ \citenamefont {Schutz}}]{Jaranowski1998}%
  \BibitemOpen
  \bibfield  {author} {\bibinfo {author} {\bibfnamefont {P.}~\bibnamefont
  {Jaranowski}}, \bibinfo {author} {\bibfnamefont {A.}~\bibnamefont {Krolak}},
  \ and\ \bibinfo {author} {\bibfnamefont {B.}~\bibnamefont {Schutz}},\ }\href
  {\doibase 10.1103/PhysRevD.58.063001} {\bibfield  {journal} {\bibinfo
  {journal} {Physical Review D}\ }\textbf {\bibinfo {volume} {58}},\ \bibinfo
  {pages} {38} (\bibinfo {year} {1998}{\natexlab{b}})},\ \Eprint
  {http://arxiv.org/abs/9804014} {arXiv:9804014 [gr-qc]} \BibitemShut {NoStop}%
\bibitem [{\citenamefont {Coughlin}\ \emph {et~al.}(2010)\citenamefont
  {Coughlin}, \citenamefont {the Ligo Scientific~Collaboration},\ and\
  \citenamefont {the Virgo~Collaboration}}]{Coughlin2010}%
  \BibitemOpen
  \bibfield  {author} {\bibinfo {author} {\bibfnamefont {M.}~\bibnamefont
  {Coughlin}}, \bibinfo {author} {\bibnamefont {the Ligo
  Scientific~Collaboration}}, \ and\ \bibinfo {author} {\bibnamefont {the
  Virgo~Collaboration}},\ }\href
  {http://stacks.iop.org/1742-6596/243/i=1/a=012010} {\bibfield  {journal}
  {\bibinfo  {journal} {Journal of Physics: Conference Series}\ }\textbf
  {\bibinfo {volume} {243}},\ \bibinfo {pages} {012010} (\bibinfo {year}
  {2010})}\BibitemShut {NoStop}%
\bibitem [{\citenamefont {Aasi}\ \emph
  {et~al.}(2015{\natexlab{b}})\citenamefont {Aasi} \emph
  {et~al.}}]{Aasi:2015mqd}%
  \BibitemOpen
  \bibfield  {author} {\bibinfo {author} {\bibfnamefont {J.}~\bibnamefont
  {Aasi}} \emph {et~al.},\ }\href {\doibase 10.1088/0264-9381/32/11/115012}
  {\bibfield  {journal} {\bibinfo  {journal} {Classical and Quantum Gravity}\
  }\textbf {\bibinfo {volume} {32}},\ \bibinfo {pages} {115012} (\bibinfo
  {year} {2015}{\natexlab{b}})}\BibitemShut {NoStop}%
\bibitem [{\citenamefont {Covas}\ \emph {et~al.}(2018)\citenamefont {Covas}
  \emph {et~al.}}]{PhysRevD.97.082002}%
  \BibitemOpen
  \bibfield  {author} {\bibinfo {author} {\bibfnamefont {P.~B.}\ \bibnamefont
  {Covas}} \emph {et~al.} (\bibinfo {collaboration} {LSC Instrument Authors}),\
  }\href {\doibase 10.1103/PhysRevD.97.082002} {\bibfield  {journal} {\bibinfo
  {journal} {Phys. Rev. D}\ }\textbf {\bibinfo {volume} {97}},\ \bibinfo
  {pages} {082002} (\bibinfo {year} {2018})}\BibitemShut {NoStop}%
\bibitem [{\citenamefont {Keitel}\ \emph {et~al.}(2014)\citenamefont {Keitel},
  \citenamefont {Prix}, \citenamefont {Papa}, \citenamefont {Leaci},\ and\
  \citenamefont {Siddiqi}}]{PhysRevD.89.064023}%
  \BibitemOpen
  \bibfield  {author} {\bibinfo {author} {\bibfnamefont {D.}~\bibnamefont
  {Keitel}}, \bibinfo {author} {\bibfnamefont {R.}~\bibnamefont {Prix}},
  \bibinfo {author} {\bibfnamefont {M.~A.}\ \bibnamefont {Papa}}, \bibinfo
  {author} {\bibfnamefont {P.}~\bibnamefont {Leaci}}, \ and\ \bibinfo {author}
  {\bibfnamefont {M.}~\bibnamefont {Siddiqi}},\ }\href {\doibase
  10.1103/PhysRevD.89.064023} {\bibfield  {journal} {\bibinfo  {journal} {Phys.
  Rev. D}\ }\textbf {\bibinfo {volume} {89}},\ \bibinfo {pages} {064023}
  (\bibinfo {year} {2014})}\BibitemShut {NoStop}%
\bibitem [{\citenamefont {Walsh}\ \emph {et~al.}(2016)\citenamefont {Walsh},
  \citenamefont {Pitkin}, \citenamefont {Oliver}, \citenamefont {D'Antonio},
  \citenamefont {Dergachev}, \citenamefont {Kr{\'{o}}lak}, \citenamefont
  {Astone}, \citenamefont {Bejger}, \citenamefont {{Di Giovanni}},
  \citenamefont {Dorosh}, \citenamefont {Frasca}, \citenamefont {Leaci},
  \citenamefont {Mastrogiovanni}, \citenamefont {Miller}, \citenamefont
  {Palomba}, \citenamefont {Papa}, \citenamefont {Piccinni}, \citenamefont
  {Riles}, \citenamefont {Sauter},\ and\ \citenamefont {Sintes}}]{Walsh2016}%
  \BibitemOpen
  \bibfield  {author} {\bibinfo {author} {\bibfnamefont {S.}~\bibnamefont
  {Walsh}}, \bibinfo {author} {\bibfnamefont {M.}~\bibnamefont {Pitkin}},
  \bibinfo {author} {\bibfnamefont {M.}~\bibnamefont {Oliver}}, \bibinfo
  {author} {\bibfnamefont {S.}~\bibnamefont {D'Antonio}}, \bibinfo {author}
  {\bibfnamefont {V.}~\bibnamefont {Dergachev}}, \bibinfo {author}
  {\bibfnamefont {A.}~\bibnamefont {Kr{\'{o}}lak}}, \bibinfo {author}
  {\bibfnamefont {P.}~\bibnamefont {Astone}}, \bibinfo {author} {\bibfnamefont
  {M.}~\bibnamefont {Bejger}}, \bibinfo {author} {\bibfnamefont
  {M.}~\bibnamefont {{Di Giovanni}}}, \bibinfo {author} {\bibfnamefont
  {O.}~\bibnamefont {Dorosh}}, \bibinfo {author} {\bibfnamefont
  {S.}~\bibnamefont {Frasca}}, \bibinfo {author} {\bibfnamefont
  {P.}~\bibnamefont {Leaci}}, \bibinfo {author} {\bibfnamefont
  {S.}~\bibnamefont {Mastrogiovanni}}, \bibinfo {author} {\bibfnamefont
  {A.}~\bibnamefont {Miller}}, \bibinfo {author} {\bibfnamefont
  {C.}~\bibnamefont {Palomba}}, \bibinfo {author} {\bibfnamefont {M.~A.}\
  \bibnamefont {Papa}}, \bibinfo {author} {\bibfnamefont {O.~J.}\ \bibnamefont
  {Piccinni}}, \bibinfo {author} {\bibfnamefont {K.}~\bibnamefont {Riles}},
  \bibinfo {author} {\bibfnamefont {O.}~\bibnamefont {Sauter}}, \ and\ \bibinfo
  {author} {\bibfnamefont {A.~M.}\ \bibnamefont {Sintes}},\ }\href {\doibase
  10.1103/PhysRevD.94.124010} {\bibfield  {journal} {\bibinfo  {journal}
  {Physical Review D}\ }\textbf {\bibinfo {volume} {94}} (\bibinfo {year}
  {2016}),\ 10.1103/PhysRevD.94.124010},\ \Eprint
  {http://arxiv.org/abs/1606.00660} {arXiv:1606.00660} \BibitemShut {NoStop}%
\bibitem [{\citenamefont {{LIGO Scientific Collaboration}}(2018)}]{lalsuite}%
  \BibitemOpen
  \bibfield  {author} {\bibinfo {author} {\bibnamefont {{LIGO Scientific
  Collaboration}}},\ }\href {\doibase 10.7935/GT1W-FZ16} {\enquote {\bibinfo
  {title} {{LIGO} {A}lgorithm {L}ibrary - {LALS}uite},}\ }\bibinfo
  {howpublished} {free software (GPL)} (\bibinfo {year} {2018})\BibitemShut
  {NoStop}%
\bibitem [{\citenamefont {Behnke}\ \emph {et~al.}(2015)\citenamefont {Behnke},
  \citenamefont {Papa},\ and\ \citenamefont {Prix}}]{Behnke2015}%
  \BibitemOpen
  \bibfield  {author} {\bibinfo {author} {\bibfnamefont {B.}~\bibnamefont
  {Behnke}}, \bibinfo {author} {\bibfnamefont {M.~A.}\ \bibnamefont {Papa}}, \
  and\ \bibinfo {author} {\bibfnamefont {R.}~\bibnamefont {Prix}},\ }\href
  {\doibase 10.1103/PhysRevD.91.064007} {\bibfield  {journal} {\bibinfo
  {journal} {Physical Review D - Particles, Fields, Gravitation and Cosmology}\
  }\textbf {\bibinfo {volume} {91}},\ \bibinfo {pages} {1} (\bibinfo {year}
  {2015})},\ \Eprint {http://arxiv.org/abs/1410.5997} {arXiv:1410.5997}
  \BibitemShut {NoStop}%
\bibitem [{\citenamefont {Prix}(2007)}]{Prix2007}%
  \BibitemOpen
  \bibfield  {author} {\bibinfo {author} {\bibfnamefont {R.}~\bibnamefont
  {Prix}},\ }\href {\doibase 10.1103/PhysRevD.75.023004} {\bibfield  {journal}
  {\bibinfo  {journal} {Physical Review D - Particles, Fields, Gravitation and
  Cosmology}\ }\textbf {\bibinfo {volume} {75}} (\bibinfo {year} {2007}),\
  10.1103/PhysRevD.75.023004},\ \Eprint {http://arxiv.org/abs/0606088}
  {arXiv:0606088 [gr-qc]} \BibitemShut {NoStop}%
\bibitem [{\citenamefont {Aasi}\ \emph
  {et~al.}(2015{\natexlab{c}})\citenamefont {Aasi} \emph {et~al.}}]{DetCharS6}%
  \BibitemOpen
  \bibfield  {author} {\bibinfo {author} {\bibfnamefont {J.}~\bibnamefont
  {Aasi}} \emph {et~al.},\ }\href
  {http://stacks.iop.org/0264-9381/32/i=11/a=115012} {\bibfield  {journal}
  {\bibinfo  {journal} {Classical and Quantum Gravity}\ }\textbf {\bibinfo
  {volume} {32}},\ \bibinfo {pages} {115012} (\bibinfo {year}
  {2015}{\natexlab{c}})}\BibitemShut {NoStop}%
\end{thebibliography}%

\end{document}